\documentclass[useAMS,usenatbib]{mn2e}
\pdfoutput=1
\usepackage{aas_macros}
\usepackage{epsfig}
\usepackage{amsmath}
\usepackage{amssymb}
\usepackage{comment}
\usepackage{caption}
\usepackage{leftidx}
\usepackage{natbib}
\usepackage[rgb]{xcolor}
\definecolor{MyGreen}{rgb}{0.0,0.6,0.3}
\definecolor{MyPurple}{rgb}{0.6,0,0.3}
\usepackage{fancyref}
\usepackage{hyperref}
\hypersetup{colorlinks=true,citecolor=MyGreen,linkcolor=MyPurple,urlcolor=blue}

\def\beq{\begin{equation}}
\def\eeq{\end{equation}}
\def\ba{\begin{eqnarray}}
\def\ea{\end{eqnarray}}
\def\bal{\begin{align}}
\def\eal{\end{align}}
\def\bxi{{\mbox{\boldmath $\xi$}}}
\def\bnab{{\mbox{\boldmath $\nabla$}}}

\begin{document}

\title[Slowing Stellar Spins] {Slowing the Spins of Stellar Cores}

\author[Fuller, Piro, \& Jermyn]{
Jim Fuller$^{1}$\thanks{Email: jfuller@caltech.edu},
Anthony L. Piro$^{2}$,
Adam S. Jermyn$^{3}$
\\$^1$TAPIR, Mailcode 350-17, California Institute of Technology, Pasadena, CA 91125, USA
\\$^2$The Observatories of the Carnegie Institution for Science, 813 Santa Barbara St., Pasadena, CA 91101, USA
\\$^3$Kavli Institute for Theoretical Physics, Kohn Hall, University of California, Santa Barbara, CA 93106, USA
}

\label{firstpage}
\maketitle

\begin{abstract}

The angular momentum (AM) evolution of stellar interiors, along with the resulting rotation rates of stellar remnants, remains poorly understood. Asteroseismic measurements of red giant stars reveal that their cores rotate much faster than their surfaces, but much slower than theoretically predicted, indicating an unidentified source of AM transport operates in their radiative cores. Motivated by this, we investigate the magnetic Tayler instability and argue that it saturates when turbulent dissipation of the perturbed magnetic field energy is equal to magnetic energy generation via winding. This leads to larger magnetic field amplitudes, more efficient AM transport, and smaller shears than predicted by the classic Tayler-Spruit dynamo. We provide prescriptions for the effective AM diffusivity and incorporate them into numerical stellar models, finding they largely reproduce (1) the nearly rigid rotation of the Sun and main sequence stars, (2) the core rotation rates of low-mass red giants during hydrogen shell and helium burning, and (3) the rotation rates of white dwarfs. We discuss implications for stellar rotational evolution, internal rotation profiles, rotational mixing, and the spins of compact objects.

\end{abstract}


\section{Introduction}

One of the longstanding problems in stellar astrophysics is the nature of angular momentum (AM) transport within evolving stars. After the main sequence, the stellar core contracts and spins up, while the envelope expands and spins down. The differential rotation may source various (magneto)-hydrodynamical instabilities that can transport AM outwards to slow the rotation of the stellar core, with crucial consequences for the spins of white dwarfs (WDs), neutron stars, and black holes. However, the AM transport mechanisms at work remain controversial and enigmatic. 

Asteroseismic observations have revolutionized this field by measuring internal stellar rotation rates for stars at various stages of evolution. Helioseismic inversions reveal nearly rigid rotation in the Sun's radiative zone \citep{howe:09,gough:15}. For low-mass ($M \! \lesssim \! 3 \, M_\odot$) stars, internal rotation rates have been measured on the main sequence \citep{kurtz:14,saio:15,benomar:15,vanreeth:18}, sub-giant/red giant branch (RGB) \citep{beck:12,mosser:12,deheuvels:14,triana:17,gehan:18}, red clump \citep{mosser:12,deheuvels:15}, and finally in WD remnants \citep{hermes:17}. The conclusion drawn from these measurements is unambiguous: core rotation rates are relatively slow, and the vast majority of AM is extracted from stellar cores as they evolve. An efficient AM transport mechanism must be at work, causing cores and compact remnants to spin orders of magnitude slower than they would in the absence of AM transport. 

In fact, the spin rates red giant cores and WDs are slower than theoretically predicted by nearly all AM transport mechanisms \citep{cantiello:14,fuller:15,belkacem:15,spada:16,eggenberger:17,ouazzani:18}. The MHD instability known as the Tayler-Spruit dynamo \citep{spruit:02} can provide more efficient AM transport than most other mechanisms, but prior implementations still predict spin rates roughly an order of magnitude too large because they struggle to overcome the steep composition gradient in red giants that suppresses AM mixing \citep{cantiello:14}. Magnetorotational instability \citep{balbus:94} may operate in some stars (e.g., \citealt{kagan:14,wheeler:15,rudiger:15}) but it is also inhibited by composition gradients and thus has difficulty operating in red giants. Another possibility is that magnetic fields enforce rigid rotation in radiative regions of stars \citep{mestel:53}, but that differential rotation develops in deep convective envelopes \citep{kissin:15}, discussed further in \autoref{disc}. See \cite{aerts:18} for a review of asteroseismic rotation rates and angular momentum transport mechanisms.

In this paper, we re-investigate the physics of the Tayler instability and its resulting saturation, as described in the seminal paper by  \citealt{spruit:02} (see also references therein \citealt{acheson:78,pitts:85,spruit:99,braithwaite:06,denissenkov:07,zahn:07}). We show that the instability can persist in RGB stars despite the existence of strong composition gradients, and we argue that its growth will saturate in a different manner than proposed by \cite{spruit:02}. In our formulation, the instability can grow to larger amplitudes and produce stronger magnetic torques. We develop a convenient prescription for the effective AM/chemical diffusivity created by the instability and implement it into stellar evolution models. The core rotation rates of these models roughly match those observed in main sequence stars, red giant cores, and WDs. Hence, if the Tayler instability operates as we propose, it may largely solve the AM transport problem in stellar interiors.

\section{Tayler Instability}

Here we analyze the onset, growth, and saturation of the the Tayler instability. We follow the heuristic description of \cite{spruit:02} and use the same notation, but we address subsequent criticism by \citet{denissenkov:07} and \citet{zahn:07}. We begin by describing the main, generally agreed upon features of the Tayler instability and summarize how this instability is typically argued to saturate via the Tayler-Spruit dynamo. We then present an alternative way of thinking about the saturation, for which we explore the corresponding AM transport.

\subsection{Instability Basics}

The instability is analyzed in a rotating frame such that the local velocity is zero, but the local shear is finite. We make a number of standard assumptions that are appropriate in the context of stellar interiors, including (1) the main background component of the magnetic field is $B_\phi$ with corresponding Alfv\`{e}n frequency $\omega_{\rm A}=B_\phi/\sqrt{4\pi \rho r^2}$, (2) the angular rotation frequency $\Omega$ is roughly constant on spherical shells since horizontal turbulence can redistribute angular momentum latitudinally much faster than it can radially, and (3) the key frequencies are ordered such that $\omega_{\rm A} \ll \Omega \ll N$, where $N$ is the Brunt-V\"ais\"al\"a frequency.

As shear winds the magnetic field, $B_\phi$ grows and becomes Tayler unstable when it reaches a critical strength of \citep{spruit:02,zahn:07}
\beq
	\label{omegac}
	\omega_{\rm A} > \omega_c \sim \Omega \bigg(\frac{N}{\Omega}\bigg)^{\! 1/2} \bigg(\frac{\eta}{r^2 \Omega}\bigg)^{\! 1/4} \, ,
\eeq
where $\eta$ is the magnetic diffusivity. The corresponding growth rate of this instability is largest for $m=1$ perturbations and is approximately
\beq
	\label{omegagrow}
	\omega_{\rm grow} \sim \frac{\omega_{\rm A}^2}{\Omega} \quad  {\rm for} \, \, \omega_{\rm A} \lesssim 2 \Omega \,.
\eeq
Due to the strong stratification in these stars, the radial length scale of the instability is limited to 
\beq
	\label{lr}
	l_r \sim \frac{1}{k_r} \lesssim l_\perp \frac{\omega_{\rm A}}{N} \, ,
\eeq
while the maximum horizontal length scale of the instability is $l_\perp \sim r$.\footnote{Although \cite{denissenkov:07} argue the instability operates on shorter length scales, we demonstrate in a forthcoming paper (Ma \& Fuller 2019, in preparation) $l_\perp \sim r$ is generally appropriate by deriving the dispersion relation at non-polar latitudes. The critique by \cite{denissenkov:07} is incorrect because it confuses the instability length scale $l_\perp \sim 1/k_\perp$ with the displacement amplitude $\xi_\perp$.}

At the short radial length scales characteristic of Tayler instability in red giants, thermal diffusion is efficient so that the thermal stratification is largely mitigated (see \autoref{appendix}). The main effect of this can be replicated by replacing $N$ in the above expressions with an ``effective'' Brunt-V\"ais\"al\"a frequency
\beq
	\label{neff}
	N_{\rm eff}^2 \simeq \frac{\eta}{K} N_T^2 + N_\mu^2 \, ,
\eeq
where $K$ is the thermal diffusivity, $N_T^2$ is the thermal component of the stratification, and $N_\mu^2$ is the compositional component. Red giant cores have large composition gradients, so $N_{\rm eff}$ in much of the core (and especially at the hydrogen-burning shell, the bottleneck for AM transport) is dominated by its compositional component, and thus $N_{\rm eff} \simeq N_\mu$. \autoref{neffs} discusses the appropriate value of $N_{\rm eff}$ when thermal diffusion is moderately important.

\subsection{Saturation via the Tayler-Spruit Dynamo}
\label{sec:ts saturation}

\begin{figure*}
\begin{center}
\includegraphics[scale=0.5]{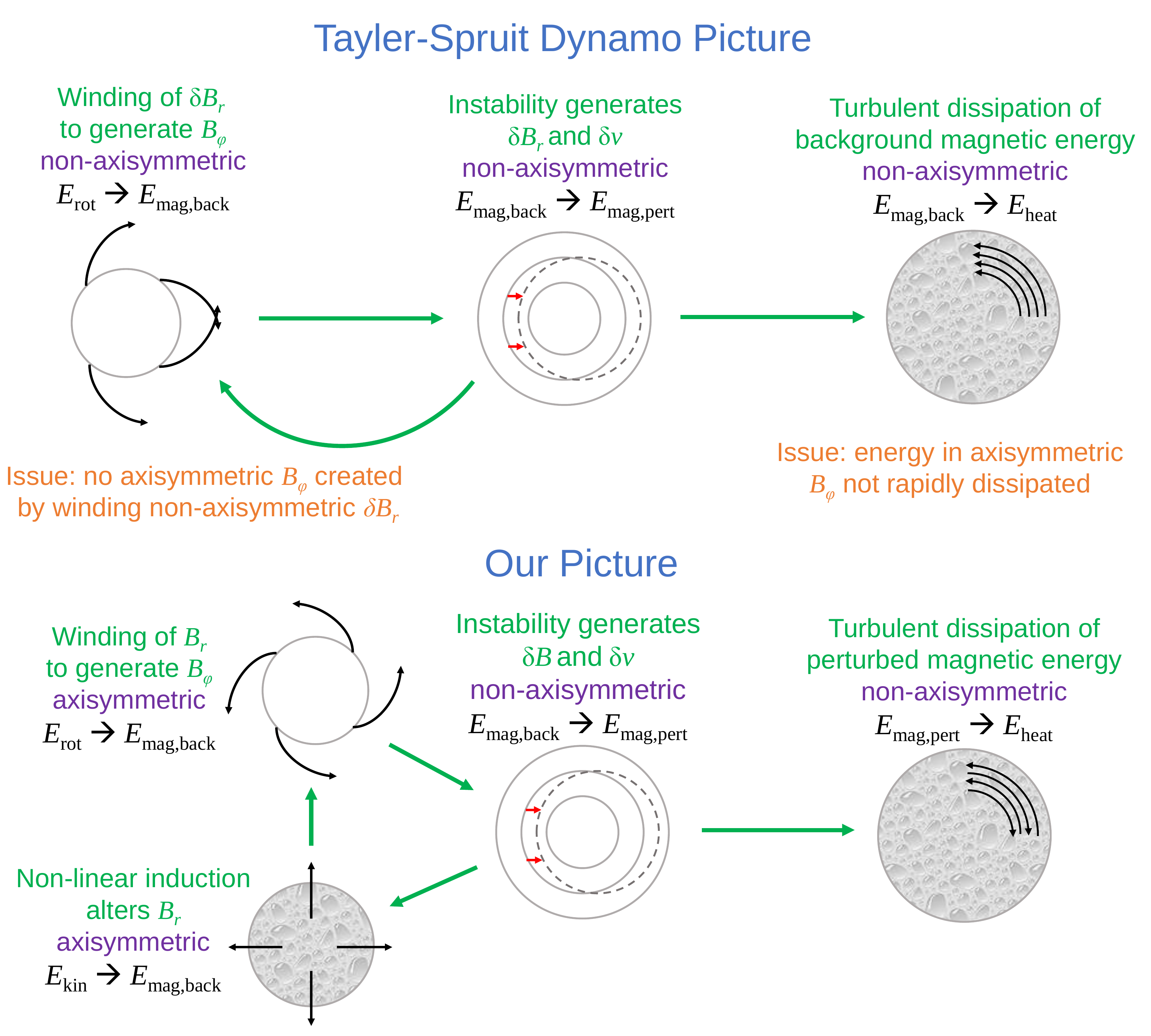}
\end{center} 
\caption{\label{Energy} Schematic showing the physical processes at work in stars undergoing Tayler instability, according to the Tayler-Spruit dynamo as proposed by \citealt{spruit:02} (top), and our model (bottom). Black arrows represent magnetic field lines, while red arrows represent fluid motions. Orange text describes processes that we argue operate differently than proposed by \citealt{spruit:02}. }
\end{figure*}

The saturation of the Tayler instability is crucial for understanding the strength of the AM transport and chemical mixing it generates.The linear instability calculation allows us to determine the rate at which energy is transferred from background fields to perturbed fields, but energy dissipation only results from non-linear effects. This non-linear energy dissipation rate is necessary for calculating the mean amplitudes of the background and perturbed fields. One possibility is that the instability grows  until it reaches a statistically stationary state in which the turbulent velocity field produces an effective viscosity or magnetic diffusivity large enough to balance the linear growth rate of the instability \citep{spruit:02}. Equating the turbulent damping rate $\gamma_{\rm turb}$ with the linear growth rate results in
\begin{equation}
	\label{gamdamp}
	\gamma_{\rm turb} \sim k_r^2 \eta_{\rm eff} \sim \frac{\omega_{\rm A}^2}{\Omega} \, ,
\end{equation}
where the wavenumber $k_r$ is the minimum required for instability, $k_r \sim \omega_{\rm A}/(N r)$, and $\eta_{\rm eff}$ is an effective turbulent diffusivity. Next, since the azimuthal field grows via winding by shear as
\beq
	\frac{\partial}{\partial t} B_\phi = q \Omega B_r \, 
\eeq
where $B_r$ is the radial field, then the amplification rate is $\gamma_{\rm amp} = q \Omega B_r/B_\phi$.  The incompressible nature of the instability implies that $k_r B_r \sim k_\perp B_\phi$ and thus $B_r \sim (\omega_{\rm A}/N) B_\phi$, such that
\beq
	\label{gamamp}
	\gamma_{\rm amp} \sim q \Omega \frac{\omega_{\rm A}}{N} \, .
\eeq
If the azimuthal field $B_\phi$ is turbulently damped at the rate of equation \ref{gamdamp}, setting equations (\ref{gamdamp}) and (\ref{gamamp}) equal determines the azimuthal field strength at saturation $B_\phi/\sqrt{4 \pi \rho r^2} \sim \omega_{\rm A} \sim q \Omega^2/N$. The radial field strength is then $B_r/\sqrt{4 \pi \rho r^2} \sim q^2 \Omega^4/N^3$, so that
\beq
	T =  B_r B_\phi \sim 4 \pi \rho r^2 \Omega^2 q^3 \bigg( \frac{\Omega}{N} \bigg)^4 \, .
\eeq
is the resulting Maxwell stress.

An important issue regarding this picture (as pointed out by \citealp{zahn:07}) is that to linear order the Tayler instability grows fastest in the non-axisymmetric $m=1$ mode. Therefore the radial field generated by the instability is non-axisymmetric, and winding of this field produces no net increase in the axisymmetric component of $B_\phi$. Thus the axisymmetric component of $B_r$ is not necessarily related to the axisymmetric component of $B_\phi$ via $B_r /B_\phi \sim \omega_{\rm A}/{N}$.

A second potential issue is that equation~(\ref{gamdamp}) may not predict the correct damping rate for a large-scale background field $B_\phi$ that varies on lengthscales much larger than $1/k_r$. If the background field $B_\phi$ is essentially constant on this lengthscale,  displacements do not mix background field lines of opposite polarity such that reconnection or dissipation occurs. Loops of background field {\it can} dissipate via reconnection if they migrate to a pole of a star where the loop has a small spatial scale, but we show in Appendix \ref{migration} that this mechanism produces a damping rate much smaller than equation \ref{gamdamp}. Hence, we believe equation \ref{gamdamp} overestimates the decay rate of any large scale component of $B_\phi$, and the saturated values of $B_r$ and $B_\phi$ can be larger than those above.

\autoref{Energy} presents a schematic for understanding the saturation of the Tayler instability as envisaged by \cite{spruit:02}, and our proposed modifications discussed below.

\subsection{Saturation via Magnetic Cascade}
\label{sec:ke saturation}

Motivated by these difficulties of calculating a turbulent/non-linear energy damping rate, we explore how turbulent cascades damp energy from the fluctuating fields $\delta \vec{B}$ and $\delta \vec{v}$. Tayler instability transfers energy from large scale magnetic fields to perturbed fields $\delta \vec{B}$ that vary on the short length scale $\sim 1/k_r$. In the linear regime, $\delta \vec{B}$ and $\delta \vec{v}$ are related to each to each other via
\beq
\label{omBperp}
\delta \vec{B} = (\vec{k} \cdot \vec{B}_\phi) \vec{\xi} \, ,
\eeq
where $\vec{\xi}$ is the Lagrangian displacement associated with the instability. Using $\delta \vec{v} \simeq - i \omega_{\rm R} \vec{\xi}$, where $\omega_{\rm R} \sim \omega_{\rm A}^2/\Omega$ is the real part of the perturbation frequency, and $\vec{k} \cdot \vec{B}_\phi \simeq k_\phi B_\phi \simeq i m B_\phi/r \sin \theta$, we have
\beq
\label{omBperp2}
\delta \vec{v} \sim \frac{\omega_{\rm A}}{\Omega} \delta \vec{v}_{\rm A} \, ,
\eeq
Here we have used $m=1$ and ignore geometric terms of order unity, and $\delta \vec{v}_{\rm A} = \delta \vec{B}/\sqrt{4 \pi \rho}$ is the perturbed Alfv\'en velocity. A similar answer can be obtained by analyzing the momentum equation
\beq
\label{mom}
\frac{\partial}{\partial t} \vec{v} + \big(\vec{v} \cdot \bnab \big)  \vec{v}
	= 2 (\vec{\Omega} \times \vec{v} ) - \frac{ \bnab P}{\rho} + \frac{ (\bnab \times \vec{B} ) \times \vec{B}}{4 \pi \rho} - \vec{g} \, .
\eeq 
The dominant forces in the horizontal direction of equation (\ref{mom}) are the Coriolis and Lorentz terms. Therefore we expect quasi-magnetogeostrophic balance, as found in rapidly rotating convective simulations by \cite{augustson:16}, such that
\beq
\label{omBperp3}
\delta v_\perp \sim \frac{\omega_{\rm A}}{\Omega} \delta v_{\rm A,\perp} \, .
\eeq
where $\delta v_\perp$ and $\delta v_{\rm A,\perp}$ are the horizontal components of the perturbed velocity and Alfv\'en velocity. Since we shall find $\omega_{\rm A} \ll \Omega$ in most stellar applications, the perturbation energy is dominated by magnetic rather than kinetic energy.

Understanding how energy cascades to small (or large) scales in MHD turbulence is tricky business. We look to \cite{goldreich:95,lithwick:03,chandran:04,lithwick:07} for guidance, though these studies did not include the effects of stratification and rotation. In Appendix \ref{damp}, we attempt to account for Coriolis and buoyancy forces on the Alfv\'enic cascade rate to smaller spatial scales, finding
\beq
\label{gammaA}
\gamma_{\rm cas} \sim \frac{\delta v_{\rm A}}{r} \, .
\eeq   
Similar to the weak Alfv\'enic turbulence described by \cite{lithwick:03}, equation \ref{gammaA} is determined by the rate at which energy is transferred to smaller scales when Tayler modes scatter off one another. We assume magnetic energy cascades from the large scales of the instability to small scales where it is damped, such that the cascade rate $\gamma_{\rm cas}$ effectively represents a turbulent damping rate of the perturbed magnetic energy. The non-linear energy dissipation rate is then
\beq
\label{edamp1}
\dot{E}_{\rm damp} \sim \frac{\delta v_{\rm A}}{r} |\delta B_\perp|^2 \, .
\eeq

We do not expect energy in the background field $B_\phi$ to be damped by a turbulent cascade to small scales. This is a key difference from \cite{spruit:02}, who uses a damping rate $\dot{E}_{\rm damp} \sim \gamma_{\rm turb} |B_\phi^2| \sim (\omega_{\rm A}^2/\Omega) |B_\phi^2|$. We believe this is unphysical because the Alfv\'enic turbulence does not cause magnetic energy in the background field to cascade to small scales, it is only the Alfv\'en waves traveling along the background field (i.e., Tayler modes) that cascade to small scales where they can be damped.

As in \cite{spruit:02}, we assume the instability saturates and reaches a statistically stationary state when the instability growth rate is matched by the turbulent damping rate such that
\beq
\label{turbdamp}
	\frac{\omega_{\rm A}^2}{\Omega} \sim \frac{\delta v_{\rm A}}{r} \, .
\eeq
Note that equation \ref{turbdamp} implies that upon saturation, the perturbed and background field are related by
\beq
\delta B_\perp \sim \frac{\omega_{\rm A}}{\Omega} B_\phi \, ,
\eeq
so that the energy damping rate is 
\beq
\label{edamp2}
\dot{E}_{\rm damp} \sim \frac{\omega_{\rm A}^4}{\Omega^3} |B_\phi|^2 \, .
\eeq

Energy in the background field {\it can }be damped if field loops can reconnect with loops of opposite polarity, which can occur sufficiently close to the pole of the star where the loops have a small spatial scale. In Appendix \ref{migration}, we show that the maximum possible energy damping rate due to this effect is
\beq
\dot{E}_{\rm damp,pole} \lesssim \frac{\omega_{\rm A}^4}{\Omega^3} B_\phi^2 \, ,
\eeq
which is less than or equal to the energy damping rate of equation \ref{edamp2}. Hence, both mechanisms may contribute to saturation of the instability, but equation \ref{edamp2} is always a good estimate of the total energy damping rate.

Next, it is useful to consider the flow of energy in this system, which is as follows.
\begin{enumerate}
\item Rotational shear energy is converted to magnetic energy by winding a radial field into a toroidal field.
\item Toroidal field energy is converted by the Tayler instability into magnetic/kinetic energy associated with the perturbed magnetic/velocity field.
\item These perturbations are  damped into heat by a turbulent cascade.
\end{enumerate}
At equilibrium, the energy input by winding must equal the turbulent energy dissipation rate. The energy input by winding is
\beq
	\frac{\partial}{\partial t} E_{\rm mag} \sim B_\phi \frac{\partial}{\partial t} B_\phi \sim q \Omega B_\phi B_r \, .
\eeq
Equating the energy input rate to energy damping rate, we have
\beq
	\label{energy}
	q \Omega B_\phi B_r \sim \frac{\omega_{\rm A}^4}{\Omega^3} |B_\phi|^2 \, .
\eeq

To solve our system, we need an estimate of $B_r/B_\phi$. In Section \ref{induct}, we argue that $B_r$ can grow until Lorentz forces nearly stabilize the plasma against the growth of the Tayler instability, such that
\beq
\label{brbphiratio}
\frac{B_r}{B_\phi} \sim \frac{\omega_{\rm A}}{N_{\rm eff}} \, .
\eeq
This is the same ratio used by \cite{spruit:02}, but it arises for different reasons. Combining this with equations (\ref{turbdamp}) and (\ref{energy}), we expect the turbulent damping to saturate the Tayler instability at
\beq
	\label{bphi1}
	\frac{B_\phi}{\sqrt{4 \pi \rho r^2}} = \omega_{\rm A} \sim \Omega \bigg( \frac{q \Omega}{N_{\rm eff}} \bigg)^{\! 1/3} \, ,
\eeq
\beq
	\label{br}
	\frac{B_r}{\sqrt{4 \pi \rho r^2}} \sim \Omega \bigg( \frac{q^2 \Omega^5}{N_{\rm eff}^5} \bigg)^{\! 1/3} \, ,
\eeq
\beq
	\label{dbperp}
	\frac{\delta B_\perp}{\sqrt{4 \pi \rho r^2}} \sim \frac{\delta v_{\rm A}}{r} \sim \Omega \bigg( \frac{q \Omega}{N_{\rm eff}} \bigg)^{\! 2/3} \, ,
\eeq
\beq
	\label{vperp}
	\frac{\delta v_\perp}{r} \sim \Omega  \frac{q \Omega}{N_{\rm eff}} \, ,
\eeq
These fields can then drive AM transport and chemical mixing as further described in \autoref{sec:am transport}.

\subsubsection{The Importance of Non-linear Induction}
\label{induct}

Before providing prescriptions that can be used for stellar evolution calculations, it is helpful to address some conceptual challenges associated with this new approach to the saturation of the Tayler instability. Initially, $B_r$ can be due to a small seed field, but as this field is converted to a toroidal field and dissipation occurs, it must be replenished. This new $B_r$ can then continue to be wound by the shear, and continue the flow of energy as outlined above. The question is how this new $B_r$ is generated and how strong can it grow.

As argued in \autoref{sec:ts saturation}, closing the loop between $B_r$ and $B_\phi$ is difficult if only linear effects are considered. This can be seen by starting with the linearized induction equation,
\beq
	\label{indlin}
	\frac{ \partial }{\partial t} \delta \vec{B} = (\delta \vec{B} \cdot \bnab) \vec{v} + (\vec{B}_\phi \cdot \bnab) \delta \vec{v} \, .
\eeq
and taking the azimuthal average, which yields
\beq
	\label{indlinav}
 	\frac{ \partial }{\partial t} \langle \delta \vec{B} \rangle = 0 \, .
\eeq
This is because the perturbed magnetic/velocity field is non-axisymmetric ($m=1$) to linear order, while the background magnetic field is axisymmetric. Equation (\ref{indlinav}) conveys the argument by \cite{zahn:07} that winding of the non-axisymmetric field cannot regenerate the axisymmetric toroidal field.

However, there can be growth of the axisymmetric radial field if we include non-linear terms in the induction equation. Perturbing the induction equation to second order and taking the azimuthal average yields
\beq
	\label{indnonlinav}
 	\frac{ \partial }{\partial t} \langle \delta \vec{B} \rangle = \langle (\delta \vec{B} \cdot \bnab) \delta \vec{v} \rangle \, .
\eeq
To order of magnitude, we thus expect
\beq
	\label{indnonlinavr}
 	\frac{ \partial }{\partial t} \langle \delta B_r \rangle \sim \frac{ \langle \delta v_\perp \delta B_\perp \rangle}{r} \, .
\eeq
Hence, we expect some growth of an axisymmetric radial field due to non-linear induction, i.e., an $\alpha$-dynamo effect.

However, as shown by \cite{braithwaite:09}, Tayler instability cannot operate if $B_r$ rises above a threshold value. This occurs if magnetic tension forces due to perturbation of the radial field are larger than magnetic pressure forces driving the instability, which can be expressed as
\beq
	r^2 k_r^2 B_r^2 \gtrsim B_\phi^2 \, . 
\eeq
If $B_r$ grows until the instability is quenched, then the maximum possible value of $B_r$ (corresponding to the longest length scale unstable disturbance $r k_r \sim N_{\rm eff}/\omega_{\rm A}$) is
\beq
	\label{brbphi}
	B_r = \frac{\omega_{\rm A}}{N_{\rm eff}} B_\phi \, .
\eeq
This is identical to the condition arising from incompressibility used by \cite{spruit:02}, but it relates the axisymmetric component of $B_r$ and $B_\phi$, whereas Spruit's relation is only valid for the non-axisymmetric component of $B_r$.

\subsection{Angular Momentum Transport}
\label{sec:am transport}

The torque via Maxwell stresses in the saturated state is found from combining equations (\ref{bphi1}) and (\ref{br}),
\beq
	\label{torque}
	T = B_r B_\phi \sim 4 \pi q \rho r^2 \Omega^2 \bigg( \frac{\Omega}{N_{\rm eff}}\bigg)^2 \,
\eeq
corresponding to an effective AM diffusivity
\beq
	\label{nuam}
	\nu_{\rm AM} = \frac{T}{4 \pi \rho q \Omega} \sim r^2 \Omega \bigg( \frac{\Omega}{N_{\rm eff}} \bigg)^2 \, .
\eeq
Although these scalings apply in the case of magnetic energy dissipation balance, it is difficult to predict the exact prefactors using these analytic arguments. We therefore parameterize our result via the saturated Alfv\'en frequency, using a dimensionless parameter $\alpha$ such that
\beq
	\label{bphi}
	\omega_{\rm A} = \alpha \Omega \bigg( \frac{q \Omega}{N_{\rm eff}} \bigg)^{1/3} \, .
\eeq
The parameterized AM diffusivity is then 
\beq
	\label{AMdiff}
	\nu_{\rm AM} = \alpha^3 r^2 \Omega \bigg( \frac{\Omega}{N_{\rm eff}} \bigg)^2 \, .
\eeq
We expect $\alpha$ of order unity, and indeed in \autoref{data} we find $\alpha \approx 1$ fits the observational data. 

Combining the instability criterion given by equation~(\ref{omegac}) with the value of $\omega_{\rm A}$ in the saturated state implies a minimum shear in order for the instability to occur and saturate as outlined above. Equating (\ref{omegac}) and (\ref{bphi}), we find
\beq
	\label{qmin}
	q_{\rm min} \sim \alpha^{-3} \bigg(\frac{N_{\rm eff}}{\Omega}\bigg)^{5/2} \bigg( \frac{\eta}{r^2 \Omega} \bigg)^{3/4} \, .
\eeq
We show in \autoref{data} that this minimum shear appears to frequently be realized in red giant stars, such that $\omega_{\rm A} \sim \omega_c$ in most of the core. In this case, the core rotation rates are set mostly by the structure of the star (i.e., profiles of $N_{\rm eff}$ and $\eta$) and are very insensitive to the initial rotation rate or prior evolution of the star.

\subsection{Energetics and Mixing}

Note that our relations at saturation imply a hierarchy of rotational, background magnetic, perturbed magnetic, and kinetic energy densities:
\begin{align}
	\label{eratio}
	& E_{\rm rot} \sim 4 \pi \rho \Omega^2 r^2 \nonumber \\
	& \gg E_{\rm mag, back} \sim B_\phi^2 \sim E_{\rm rot} \bigg( \frac{q \Omega}{N_{\rm eff}} \bigg)^{\! 2/3} \nonumber \\
	& \gg E_{\rm mag, pert} \sim \big|\delta B|^2 \sim E_{\rm rot} \bigg( \frac{q \Omega}{N_{\rm eff}} \bigg)^{\! 4/3} \nonumber \\
	& \gg E_{\rm kin} \sim 4 \pi \rho |\delta v|^2 \sim  E_{\rm rot} \bigg( \frac{q \Omega}{N_{\rm eff}} \bigg)^{\! 2} \, .
\end{align}
These hierarchies are true as long as $q \lesssim N_{\rm eff}/\Omega$, which is true in our models in \autoref{data} where $q \sim 1$ and $N_{\rm eff}/\Omega \sim 10^4$. However, in cases where $q$ is much larger the hierarchy will be altered, and the instability could saturate in a different manner. 

From the divergence-free conditions on the perturbed magnetic and velocity fields, we can also calculate their radial components:
\beq
\label{dBr}
\frac{\delta B_r}{\sqrt{4 \pi \rho r^2}} \simeq \frac{k_\perp}{k_r} \frac{\delta B_\perp}{\sqrt{4 \pi \rho r^2}} \sim \frac{\omega_{\rm A}}{N_{\rm eff}} \frac{\delta B_\perp}{\sqrt{4 \pi \rho r^2}} \sim \Omega \frac{ q \Omega^2}{N_{\rm eff}^2} \, ,
\eeq
\beq
\label{dvr}
\frac{\delta v_r}{r} \sim \frac{\omega_{\rm A}}{N_{\rm eff}} \frac{\delta v_\perp}{r} \sim \Omega \bigg( \frac{ q^4 \Omega^7}{N_{\rm eff}^7} \bigg)^{\! 1/3} \, .
\eeq
The radial components of the fields are typically orders of magnitude smaller than the horizontal components due to the tiny value of $\Omega/N_{\rm eff}$ in most stars. 

For this reason, chemical mixing induced by the Tayler instability will likely be less important than AM transport in most stars. The effective chemical mixing diffusivity is
\beq
	\label{diffr}
	\nu_{\rm mix} \sim \delta v_r l_r \,.
\eeq
Using the relations above, we have 
\beq
	\label{diffr2}
	\nu_{\rm mix} \sim r^2 \Omega
	 \bigg( \frac{\Omega}{N_{\rm eff}} \bigg)^2
	 \bigg(\frac{q \Omega}{N_{\rm eff}}\bigg)^{5/3} \, ,
\eeq
so that
\beq
	\label{diffrat}
	\frac{\nu_{\rm mix}}{\nu_{\rm AM}} \sim  \bigg(\frac{q \Omega}{N_{\rm eff}}\bigg)^{5/3} \, ,
\eeq
for the ratio of chemical mixing to AM transport.

In red giants, we find $\nu_{\rm mix}/\nu_{\rm AM} \sim 10^{-6}$, such that chemical mixing caused by the Tayler instability is minuscule. The chemical mixing timescale across the star is longer the Ohmic diffusion time scale, which is longer than the stellar evolution timescale \citep{cantiello:16}, so the chemical mixing is likely negligible. The scaling of \ref{diffrat} is stronger than that of Eddington-Sweet circulation, so we expect chemical mixing from Tayler instabilities to be unimportant unless $q \gg 1$ or $N_{\rm eff} \ll N$.

\section{Stellar Models}
\label{data}

With our prediction for AM transport due to Tayler torques, we implement this prescription into stellar evolutionary models to predict their internal rotation rates. We then compare with asteroseismic measurements of internal rotation rates, finding generally good agreement.

\subsection{Properties of Red Giant Cores}
\label{appendix}

\begin{figure}
\begin{center}
\includegraphics[scale=0.35]{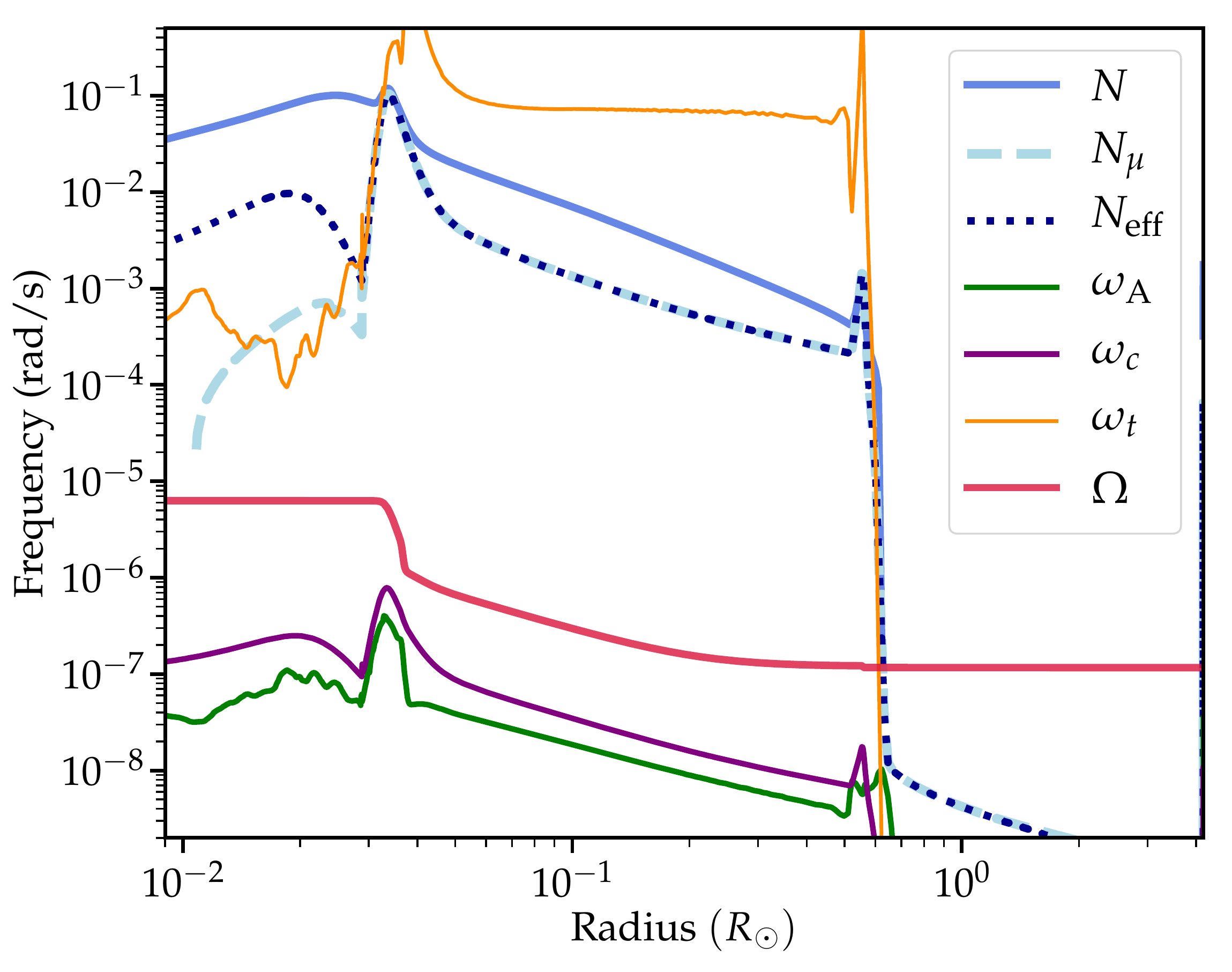}
\end{center} 
\caption{ \label{fig:model} Important frequencies as a function of radius in a $1.2 \, M_\odot$, $4 \, R_\odot$ model at the base of the RGB. We show the Brunt-V\"ais\"al\"a frequency $N$, its compositional component $N_\mu$, and its effective value $N_{\rm eff}$ (equation \ref{neff}) when thermal diffusion is important. We also plot the angular rotation frequency $\Omega$, the saturated Alfv\'en frequency $\omega_{\rm A}$ (equation \ref{bphi}), the minimum Alfv\'en frequency required for Tayler instability $\omega_c$ (equation \ref{omegac}), and the thermal diffusion frequency at the instability length scale $\omega_t$ (equation \ref{omegat}). Note that $\omega_{\rm A} \ll \Omega$ throughout the interior such that Tayler instability occurs in the rapidly rotating regime. Because $\omega_t \gg \omega_{\rm A}$, the instability occurs in the limit where thermal diffusion is important.}
\end{figure}

An important feature of post-main sequence stars is their steep composition gradient in and above their hydrogen burning shells. \autoref{fig:model} shows a $M=1.2 \, M_\odot$ model on the lower RGB at $R=4.1 \, R_\odot$ and log(g)$=3.3$. At the hydrogen burning shell, the stabilization is  primarily due to the hydrogen-helium composition gradient such that $N \simeq N_\mu$, but even above the burning shell, we often find $N_\mu \sim N/5$ due to the hydrogen gradient left behind by partial pp-chain burning during the main sequence. Hence, the compositional part of the stratification is very important, even above the burning shell.

An important consideration for the operation of the Tayler instability is whether thermal diffusion will  undermine the thermal component of $N^2$. It is useful to compare the growth rate of the instability with the thermal diffusion time scale at the instability lengthscale,
\begin{align}
\label{omegat}
\gamma &= k_r^2 \chi \newline \\
& \simeq \frac{ \chi}{r^2} \frac{N_{\rm eff}^2}{\omega_{\rm A}^2} \, .
\end{align}
Here, we have used the maximum radial lengthscale for Tayler instability $l_r \sim r (\omega_{\rm A}/N_{\rm eff})$, and the thermal diffusivity $\chi = 16 \sigma_{\rm SB} T^3/(3 \rho^2 c_v \kappa)$. Thermal diffusion strongly reduces the effective thermal stratification when $\gamma \gtrsim \omega$, where $\omega$ is the real part of the frequency of the overstable oscillations. \cite{zahn:07} show that $\omega \sim \omega_{\rm A}^2/\Omega$. Using our saturated field strength (equation \ref{bphi}), we find thermal diffusion is very important when
\beq
\label{omegatcomp}
\omega_t = \frac{\chi}{r^2} \bigg( \frac{N_{\rm eff}^5}{q^2 \alpha^6 \Omega^5} \bigg)^{2/3} \gtrsim \Omega .
\eeq
A comparison of $\omega_t$ and $\Omega$ in \autoref{fig:model} shows that the former is larger throughout the radiative core, such that thermal diffusion is very important. This is almost always the case in our post-main sequence models. In this case, as discussed by \cite{zahn:07}, the effective stabilization is given by equation \ref{neff}. In most regions of our models, we find $N_{\rm eff}^2 \simeq N_\mu^2$.

\subsection{Comparison with Measurements}

We expect the AM diffusivity of equation (\ref{AMdiff}) to capture the scaling of magnetic torques in differentially rotating stars, but we must still evaluate the appropriate value of $\alpha$. To that end, we construct rotating stellar models using the MESA stellar evolution code \citep{paxton:11,paxton:13,paxton:15,paxton:18}. We assume rotation constant on spherical shells and near rigid rotation in convective zones. These models include AM transport via the diffusivity of equation (\ref{AMdiff}) applied to gradients in angular rotation frequency, which is included if equation (\ref{qmin}) is satisfied. Our models also include hydrodynamic AM transport mechanisms (which are usually negligible compared to our revised TS torques), but we do not use MESA's default prescription for TS torques. A full inlist can be found in Appendix \ref{appendixmodel}.

\begin{figure}
\begin{center}
\includegraphics[scale=0.31]{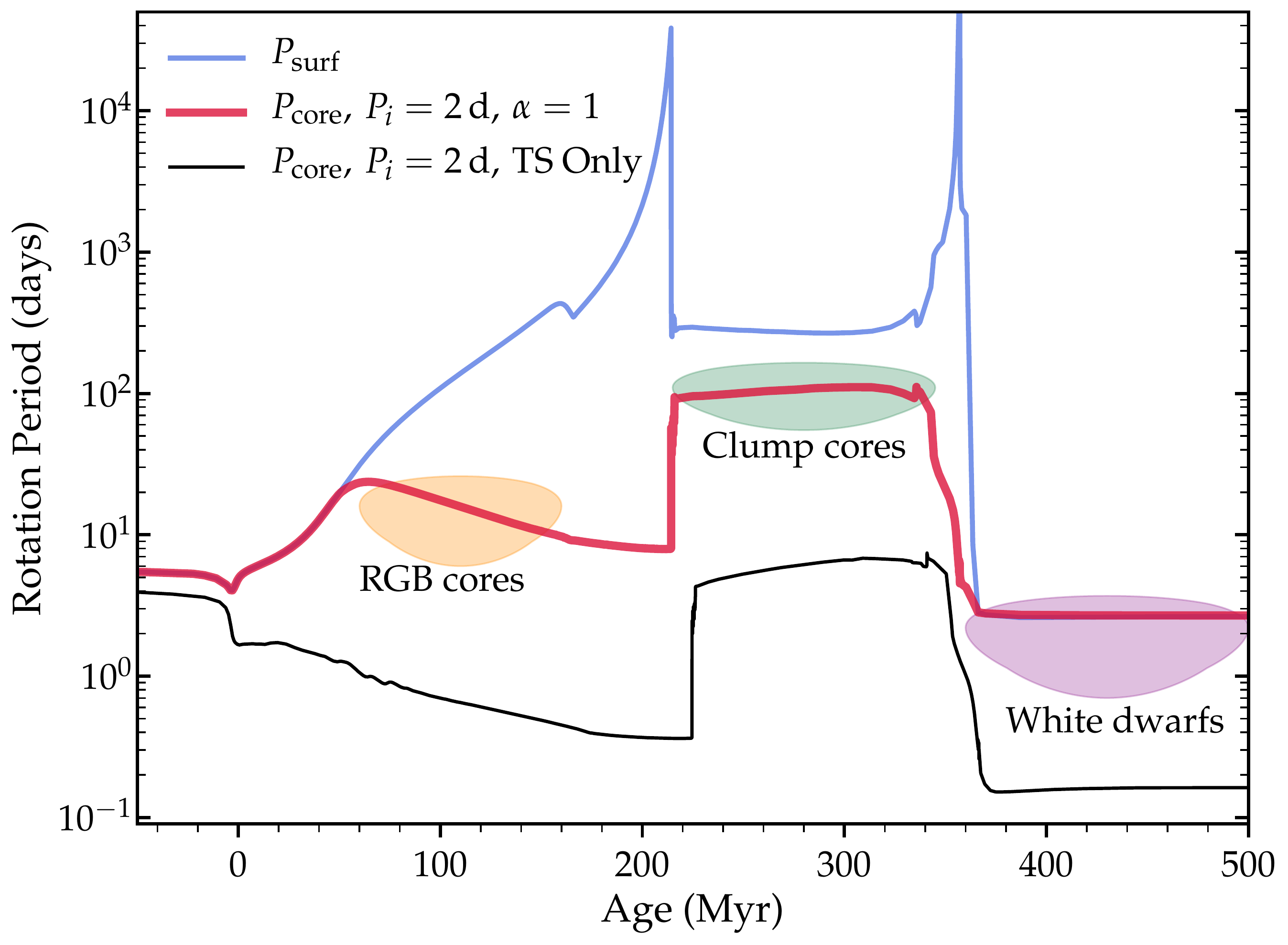}
\end{center} 
\caption{\label{fig:MRI1p6} Post-main sequence rotational evolution of a $1.6 \, M_\odot$ star, with a ZAMS rotation rate $P_i = 2$ days and including our updated prescription for AM transport with $\alpha=1$. We plot the surface rotation rate (blue line), and core rotation rate as sensed by mixed modes (red line). We also include the core rotation rate using a prior prescription for the TS dynamo (black line). Our AM transport scheme closely matches observations along the red giant branch \citep{mosser:12,gehan:18}, red clump \citep{mosser:12,deheuvels:15}, and white dwarf phase \citep{hermes:17}.
}
\end{figure}

To calculate core rotation rates $P_{\rm core}$ for comparison with asteroseismology, we compute the average core rotation period $P_{\rm core} = 2 \pi/\Omega_{\rm core}$ as sensed by a gravity wave in the WKB limit,
\beq
\label{pcore}
\Omega_{\rm core} = \frac{\int \Omega N dr/r }{\int N dr/r } \, .
\eeq
The bounds of the integral in equation \ref{pcore} correspond to the boundaries of the core gravity mode cavity where $\omega_{\rm g} \! < \! N$, and we consider gravity waves with frequency $\omega_{\rm g} = 2 \pi \nu_{\rm max}$, where the frequency of maximum power is calculated via classical scaling relations $\nu_{\rm max} = 3090 \, \mu {\rm Hz} \, (M/M_\odot) (R/R_\odot)^{-2} (T/T_\odot)^{-1/2}$.  For our WD models, we simply set $P_{\rm core}$ equal to the central rotation rate.

We evolve solar metallicity models ranging from $1.2-6 \, M_\odot$ from the zero-age main sequence (ZAMS) to the WD phase. We initiated each model with a spin rate $P_i = 2$ days, except for the $1.2 \, M_\odot$ model for which we used $P_i = 20$ days to account for main sequence magnetic braking. \autoref{fig:MRI1p6} shows evolution of the core and surface rotation rates of a $1.6 \, M_\odot$ model with a ZAMS rotation period $P_{\rm i} = 2$ days. We also denote typical measured rotation rates of cores of stars ascending the RGB \citep{mosser:12,gehan:18}, stars on the red clump \citep{mosser:12}, and WDs \citep{hermes:17}, all of which descended predominantly from main sequence stars in the range $1 \, M_\odot \lesssim M \lesssim 3 \, M_\odot$. Typical core rotation rates are $P_{\rm core} \sim 10-20$ days on the lower RGB, $P_{\rm core} \sim 50-200$ days on the red clump, and $P \sim 0.5-4$ days for WDs. 

Our models generally exhibit very similar rotation rates to observations for $\alpha\approx 1$, a reasonable value since we expect $\alpha \sim 1$. The agreement is very good along the RGB, red clump, and in the WD phase. We also plot the model's surface rotation rate, which shows that nearly rigid rotation is maintained  beyond the end of the main sequence. The rigid rotation is maintained until the ratio of $\Omega/N_{\rm eff}$ becomes sufficiently small that the Tayler instability cannot fully prevent the spin-up of the core. After this point, differential rotation develops during the late sub-giant/early red giant phase as the core contracts and tries to spin up, while the envelope expands and spins down. However, the AM transport is strong enough that the core rotation period actually increases between the main sequence and the tip of the RGB. In contrast, a model including the default prescription for TS torques exhibits core spin-up along the RGB, spinning an order of magnitude too fast compared to observations, in agreement with the results of \cite{cantiello:14}. Models with only hydrodynamic prescriptions for AM transport have even faster core rotation and are totally incompatible with observations.

Our models diverge from those with different AM transport prescriptions along the sub-giant branch and lower RGB because most of the core AM extraction in our models occurs during these phases. At later stages of evolution of low-mass stars ($M\lesssim 2 \, M_\odot$), the stabilizing composition gradients are so large in comparison to the local rotation rates that very little AM transport occurs after the RGB bump. This result agrees with \cite{cantiello:14}, who find that red clump and WD rotation rates require approximate conservation of core AM after the RGB bump. 

\begin{figure}
\begin{center}
\includegraphics[scale=0.32]{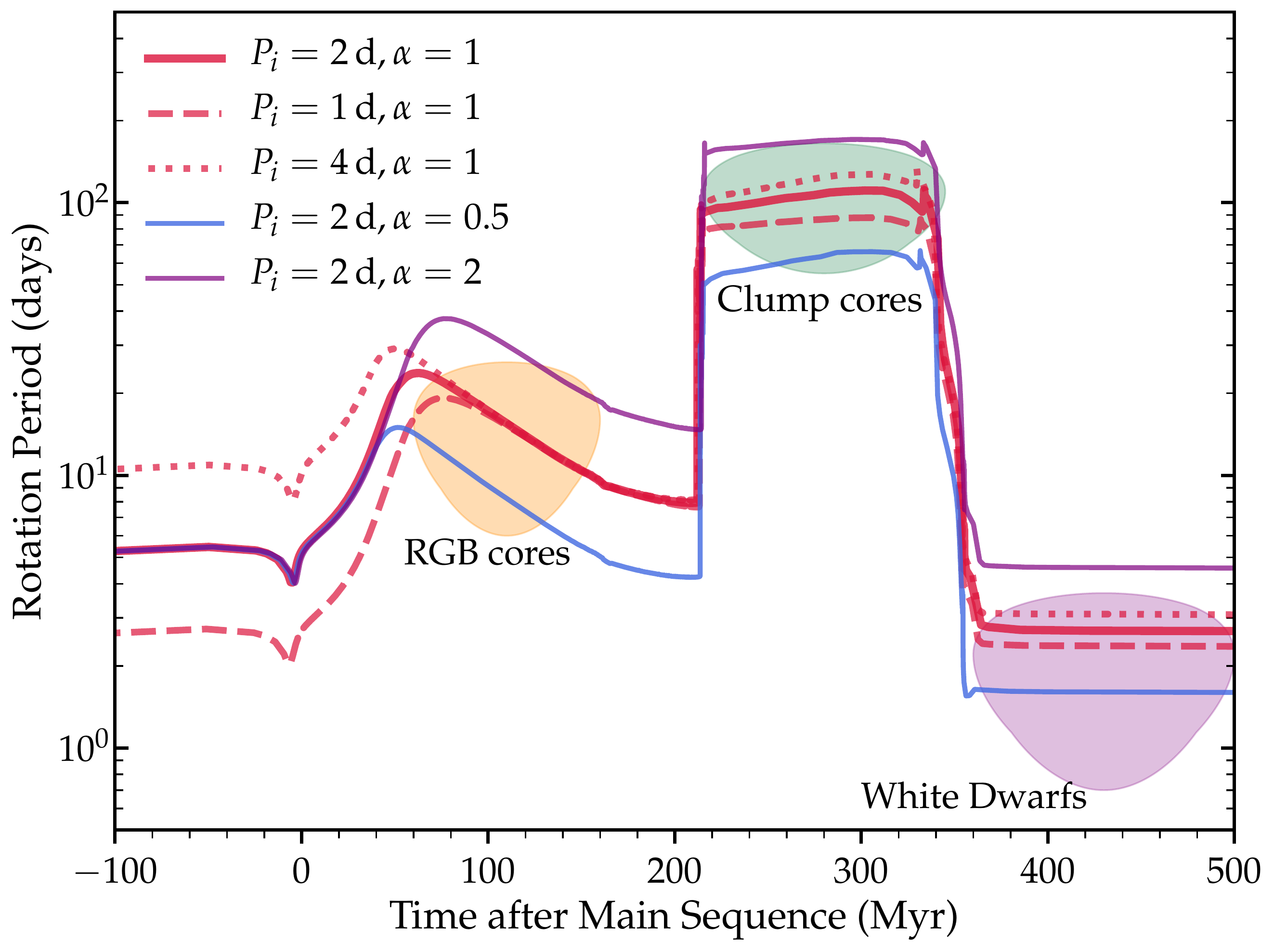}
\end{center} 
\caption{\label{fig:MRI1p6rotcomp} Core rotation rate for the same star as \autoref{fig:MRI1p6}, but varying the initial rotation rate and parameter $\alpha$. The post-main sequence rotation period scales approximately as $\alpha$, but is relatively insensitive to initial rotation rate.
}
\end{figure}

To understand the dependence of our results on the parameter $\alpha$ and the star's initial spin rate, \autoref{fig:MRI1p6rotcomp} shows core rotation rates for models with different values of $\alpha$ and $P_i$. Remarkably, the core spin rate on the RGB and red clump is nearly independent of the initial spin. The reason is the strong dependence of AM transport on the local spin rate, with $\nu_{\rm AM} \propto \Omega^3$. Rapidly rotating cores experience a stronger spin-down torque while slowly rotating cores feel a weaker spin-down torque, causing convergent migration in the post-main sequence core spin rate. Note that the WD spin rate does exhibit some dependence on initial spin rate, largely because this determines the AM of the accreted material on the clump and asymptotic giant branch (AGB). 

\autoref{fig:MRI1p6rotcomp} shows that the post-main sequence spin period is roughly proportional to $\alpha$. Nearly rigid rotation is maintained along the main sequence, regardless of $\alpha$, except for very slow rotators in which the value of $\nu_{\rm AM}$ is much smaller (see discussion in \autoref{disc}). The models slightly diverge from one another on the lower RGB, with smaller values of $\alpha$ allowing faster core rotation. We find the main effect of the value of $\alpha$ in our models is not the prefactor in equation \ref{AMdiff}, but rather in determining the minimum shear $q_{\rm min}$ (equation \ref{qmin}) required for Tayler instability to saturate as we have outlined. When $q > q_{\rm min}$, efficient AM transport generally decreases the core rotation and shear, thereby reducing $q$ until $q \sim q_{\rm min}$, as shown in \autoref{appendixmodel}. In this limit, equation \ref{qmin} predicts that the core rotation scales as $\Omega \propto \alpha^{-12/13}$, in line with our numerical finding that the core spin period is approximately proportional to $\alpha$.

\begin{figure}
\begin{center}
\includegraphics[scale=0.34]{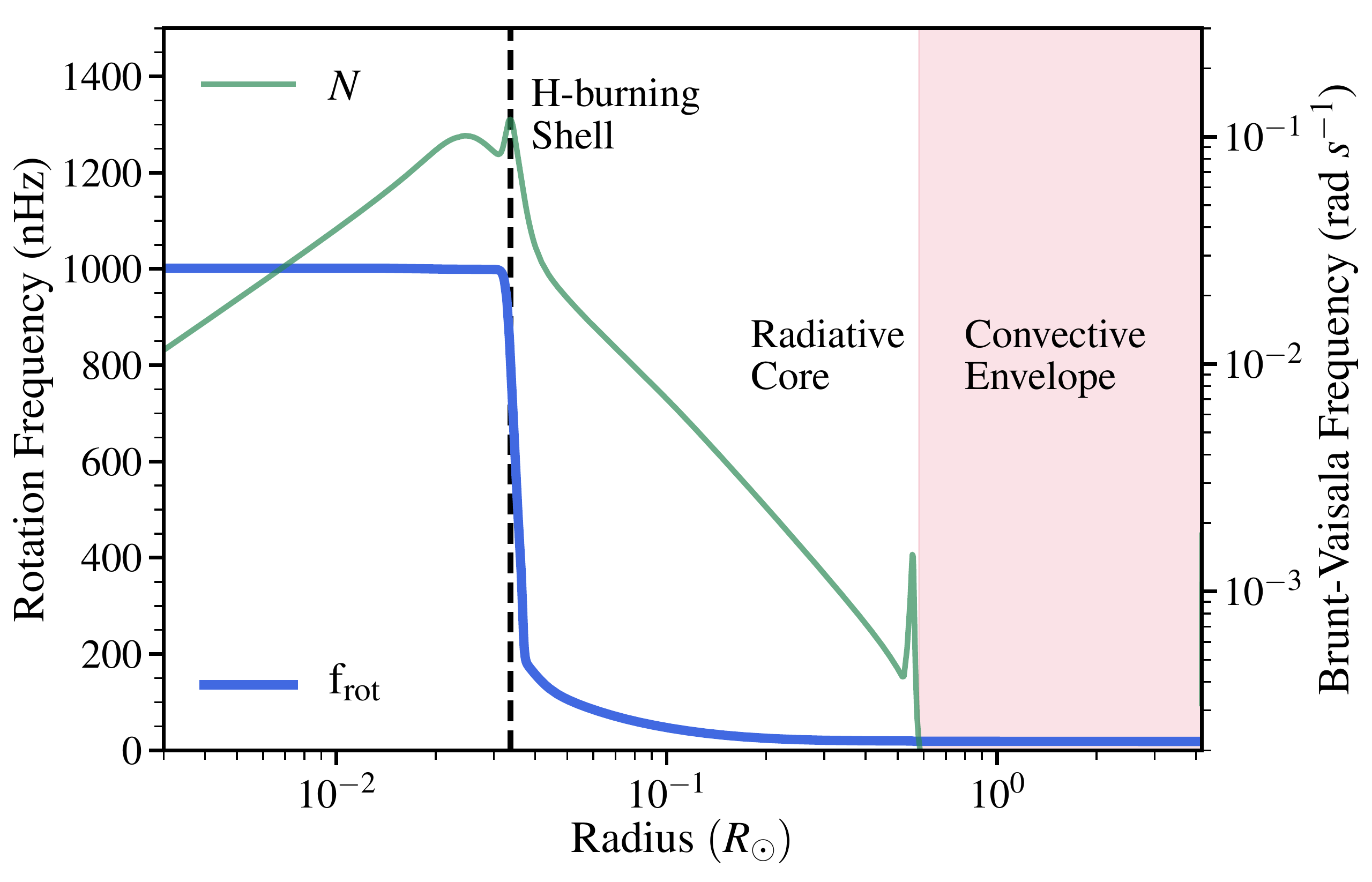}
\end{center} 
\caption{ \label{fig:MRI1p2rot_profile} Rotation profile of a $1.2 \, M_\odot$ model on the lower RGB when its radius is $R \approx 4 \, R_\odot$ (same model as Figure \ref{fig:model}). The model has $P_i = 20 \, {\rm d}$ and $\alpha=1$. The right axis shows the Brunt-V\"ais\"al\"a frequency. Shear is concentrated around the hydrogen burning shell (dashed black line) where the compositional component of the stratification is largest.
}
\end{figure}

\autoref{fig:MRI1p2rot_profile} shows the rotation profile of a $1.2 \, M_\odot$ model at the base of the RGB for $\alpha = 1$. The shear is strongest at the hydrogen burning shell where $N_\mu$ is largest. There is very little shear in the helium core where almost no compositional stratification exists. Significant shear also exists in the radiative region above the burning shell due to the composition gradient left over from incomplete hydrogen burning outside the central core during the main sequence. Encouragingly, this rotational profile is very similar to that inferred for the RGB star KIC 4448777 (\citealt{dimauro:16}, Figure 11) at nearly the same phase of evolution, though we caution that the actual rotation profile is poorly constrained by asteroseismic data.

\begin{figure}
\begin{center}
\includegraphics[scale=0.31]{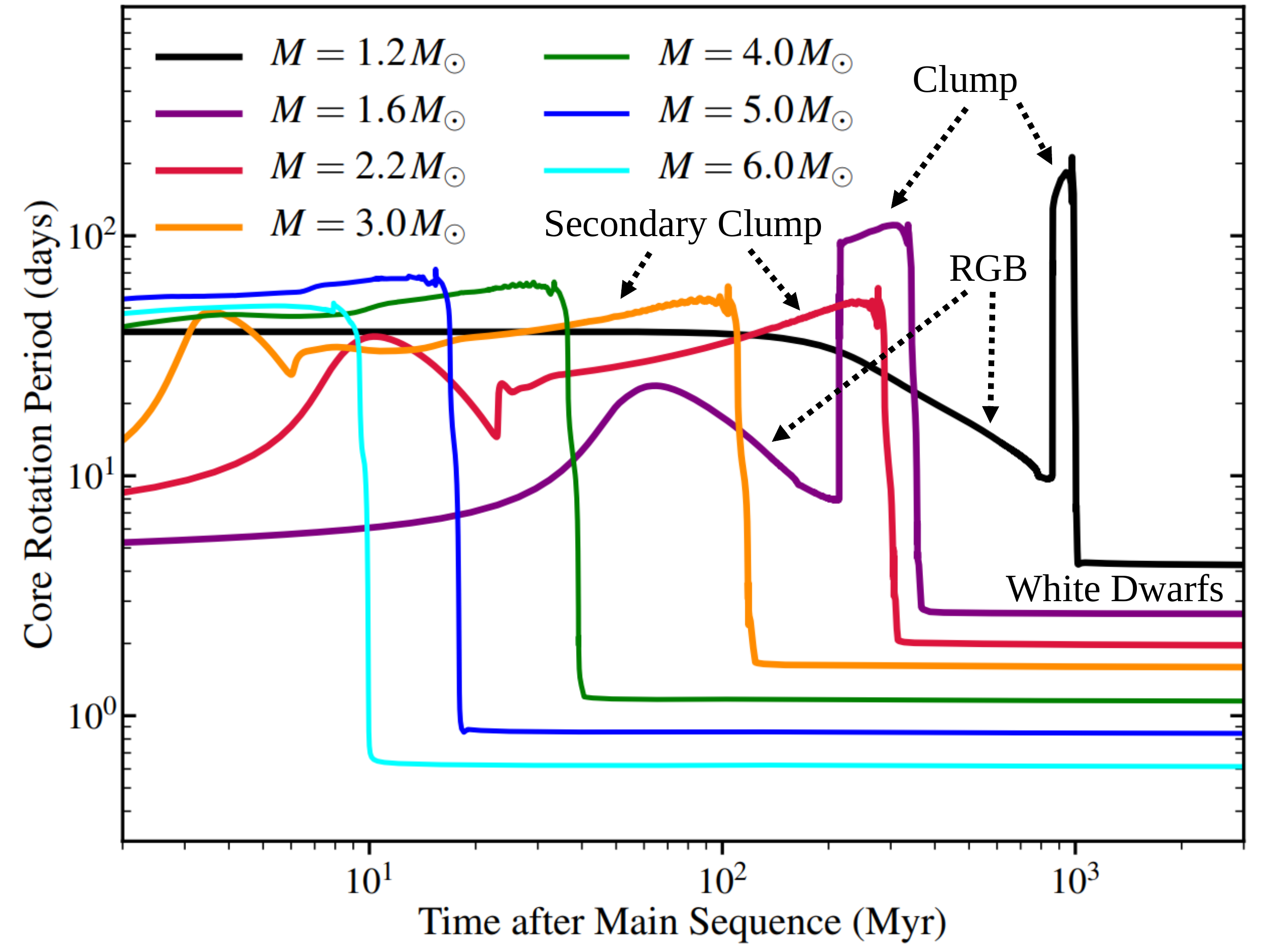}
\end{center} 
\caption{ \label{fig:MRIall} Core rotational evolution (as sensed by mixed modes) for models of several masses. Each model uses $\alpha=1$ and is initiated with $P_i = 2$ days, except the $1.2 \, M_\odot$ model which has $P_i = 20$ days. The highest plateaus of each model correspond to the core helium-burning phase, and each model ends as a carbon-oxygen white dwarf.
}
\end{figure}



Our model makes important predictions for core rotation rates as a function of progenitor mass, as shown in \autoref{fig:MRIall}. We find core rotation rates on the RGB in the range $10 \, {\rm days} \! \lesssim P_{\rm core} \! \lesssim \! 30 \, {\rm days}$ regardless of mass. On the clump, the differences between stars of different masses are slightly larger: our models predict slower core rotation rates for low-mass clump stars, with $P_{\rm core} \sim 100-200 \, {\rm days}$ for $\approx 1.2 \, M_\odot$ stars. We predict faster rotation for secondary clump stars, with $P_{\rm core} \sim 50$ days for $M \approx 2.2 \, M_\odot$. The trend of faster rotation for higher-mass clump/secondary clump stars indeed appears to be present in the results of \cite{mosser:12}. We also predict very mild core spin-down during helium-burning for lower-mass stars, whereas we predict significant core spin-down during the helium-burning phase of secondary clump stars (by a factor of $\sim \! 2$).

\begin{figure}
\begin{center}
\includegraphics[scale=0.38]{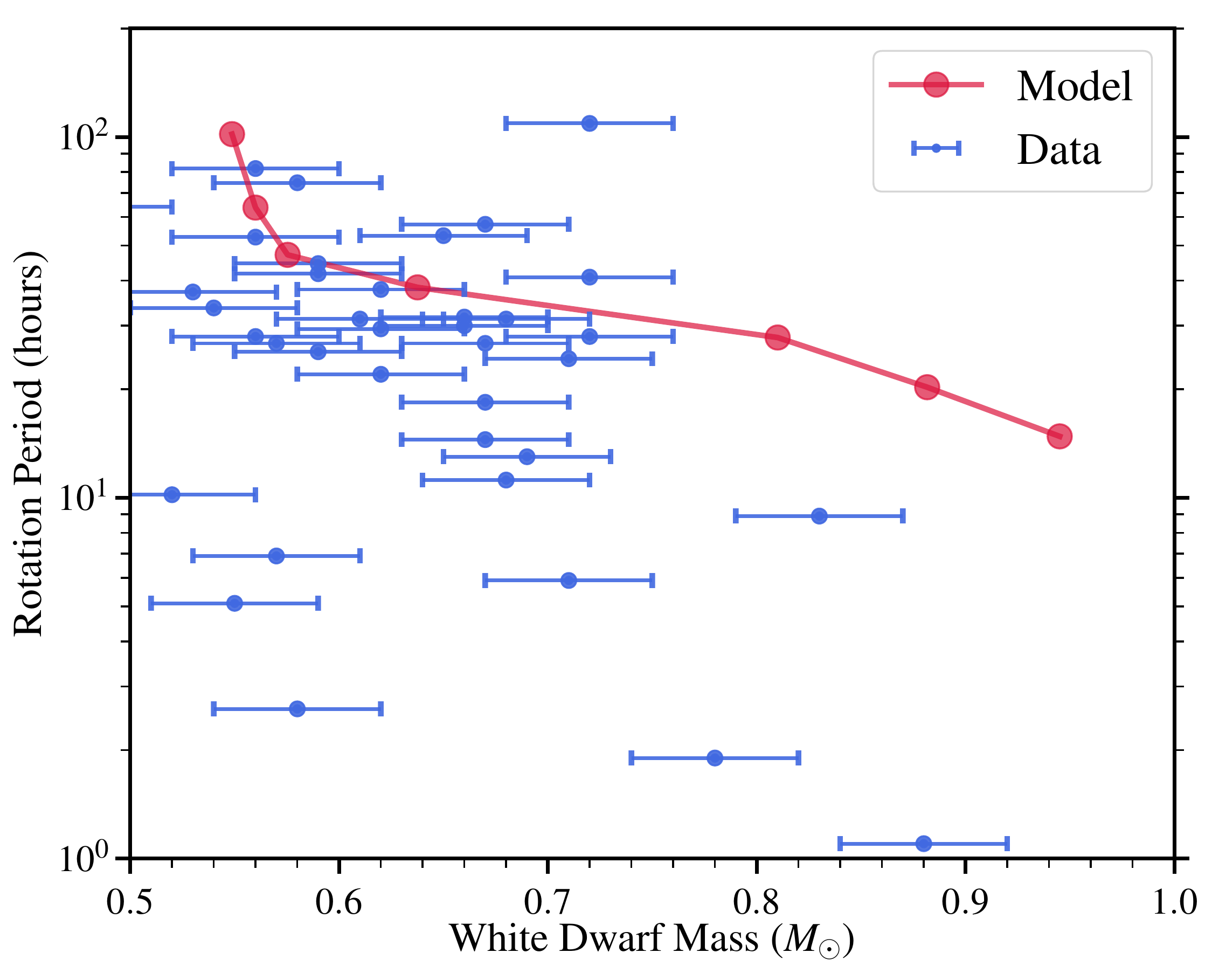}
\end{center} 
\caption{ \label{fig:MRIWDrot} Internal rotation rates of our white dwarf models as a function of white dwarf mass. The models are the same as the end points of the models in \autoref{fig:MRIall}. We also plot white dwarfs with asteroseismic rotation rates from \citealt{hermes:17}.
}
\end{figure}

We also make predictions for WD spin rates as a function of WD mass.  \autoref{fig:MRIWDrot} shows asteroseismicly measured WD spin periods from \cite{hermes:17} as a function of WD mass, along with predictions from our stellar models evolved down the WD cooling track to the ZZ-ceti instability strip. The ZAMS models have masses $M_{\rm ZAMS} =(1.2,1.6,2.2,3.0,4.0,5.0,6.0) \, M_\odot$ and produce carbon-oxygen WDs with masses $M_{\rm WD} =(0.54,0.56,0.58,0.64,0.81,0.87,0.95) \, M_\odot$.  \autoref{fig:MRIWDrot} demonstrates that our predicted WD spin rates are very similar to those observed, with $P_{\rm WD} \sim 1-3$ days for WDs with $M\approx 0.6 \, M_\odot$. W predict that more massive WDs rotate faster, a trend indeed observed in \cite{hermes:17}, but our highest mass models rotate much slower than the observed high-mass ($M \gtrsim 0.7 \, M_\odot$) WDs. 

In general, the observed population on WDs appears to exhibit more scatter than our model predictions, some of which may be inherited from the scatter in progenitor rotation rate as shown in \autoref{fig:MRI1p6rotcomp}. Additionally, our models do not take into account binary effects such as mergers (either during stellar evolution or WD mergers) that may produce faster rotating stellar cores and WDs. \cite{kilic:18} suggest that $\sim \! 10 \%$ of WDs, especially higher mass WDs, are likely to be merger products. We speculate some of the faster rotating WDs shown in \autoref{fig:MRIWDrot} resulted from stellar mergers during post-main sequence evolution. Finally, our models do not make reliable predictions for descendants of magnetic Ap/Bp stars, whose strong internal fields likely increase AM transport and may keep their cores more slowly rotating than our predictions.

\section{Discussion}
\label{disc}

Our AM transport prescription predicts extremely short AM transport times $t_{\rm AM}$ within radiative zones of main sequence stars, 
\beq
\label{tam}
t_{AM} \sim \frac{r^2}{\nu_{\rm AM}} \sim \frac{N_{\rm eff}^2}{\alpha^{3}  \Omega^3} \, .
\eeq
For a fast rotating young Sun, we find that the instability occurs in the nearly adiabatic limit, such that $N_{\rm eff} \approx N$. Evaluating equation (\ref{tam}) in the radiative zone of a young solar model rotating at $P = 3$ days yields a typical AM transport timescale $t_{\rm AM} \sim 10$ years. In the current Sun, the rotation rate is much slower and $t_{\rm AM} \sim 10^4 \, {\rm years}$, but this still enforces nearly rigid rotation in agreement with helioseismic measurements \citep{howe:09,gough:15}. In most cases, we predict nearly rigid rotation for main sequence stars, although modest differential rotation may exist in very slowly rotating stars. For rotation rates of about 100 days, equation (\ref{tam}) predicts $t_{\rm AM} \sim 10^6 \, {\rm yr}$, which may be longer than the timescale for shear to develop due to other effects such as internal gravity waves \citep{rogers:13,fullerwave:14,townsend:18}. Hence, differential rotation can persist in slowly rotating stars, and this could explain why some very slowly rotating stars (see \citealt{kurtz:14,saio:15,triana:15,kallinger:17,sowicka:17}) appear to exhibit some degree of differential rotation, while more rapidly rotating stars main sequence stars appear to be nearly rigidly rotating \citep{aerts:18}.

The short AM transport time for main sequence stars may seemingly contradict observations of rotational evolution of young $\approx \! 1.0 \, M_\odot$ stars, for which several works (e.g., \citealt{denissenkov:10,gallet:15,lanzafame:15}) find evidence for core-envelope coupling times in the range $t_{\rm AM} \! \sim \! 10-100$ Myr. These coupling times are deduced by fitting models including magnetic braking and core-envelope decoupling to the surface rotational evolution of cluster stars at a variety of ages. The wide distribution of surface rotation rates extending to ages older than $100$ Myr can only be fit using a value of $t_{\rm AM} \sim 10^7-10^8 \, {\rm yr}$ that varies with mass and rotation rate. These models all utilize relatively simple and deterministic magnetic braking laws, but the bimodal rotation rates of low-mass cluster stars (e.g., \citealt{rebull:16,rebull:17,rebull:18}) cannot be explained by such models. Instead, it appears that magnetic braking is strongly influenced by surface magnetic field morphology, such that rapidly rotating stars with more complex fields can spin down more gradually than slower rotating stars with mostly dipolar fields, which can explain the bimodality and rotational evolution of low-mass cluster stars \citep{brown:14,garraffo:15,garraffo:16,garraffo:18}, even assuming rigid internal rotation. In light of the relatively short AM transport time scales needed to explain the slow rotation of red giant cores, we find it most likely that $t_{\rm AM}$ is indeed very small for main sequence stars such that they rotate nearly rigidly, but that magnetic braking can be a more complex process than previously assumed, especially for young stars.

\begin{figure}
\begin{center}
\includegraphics[scale=0.38]{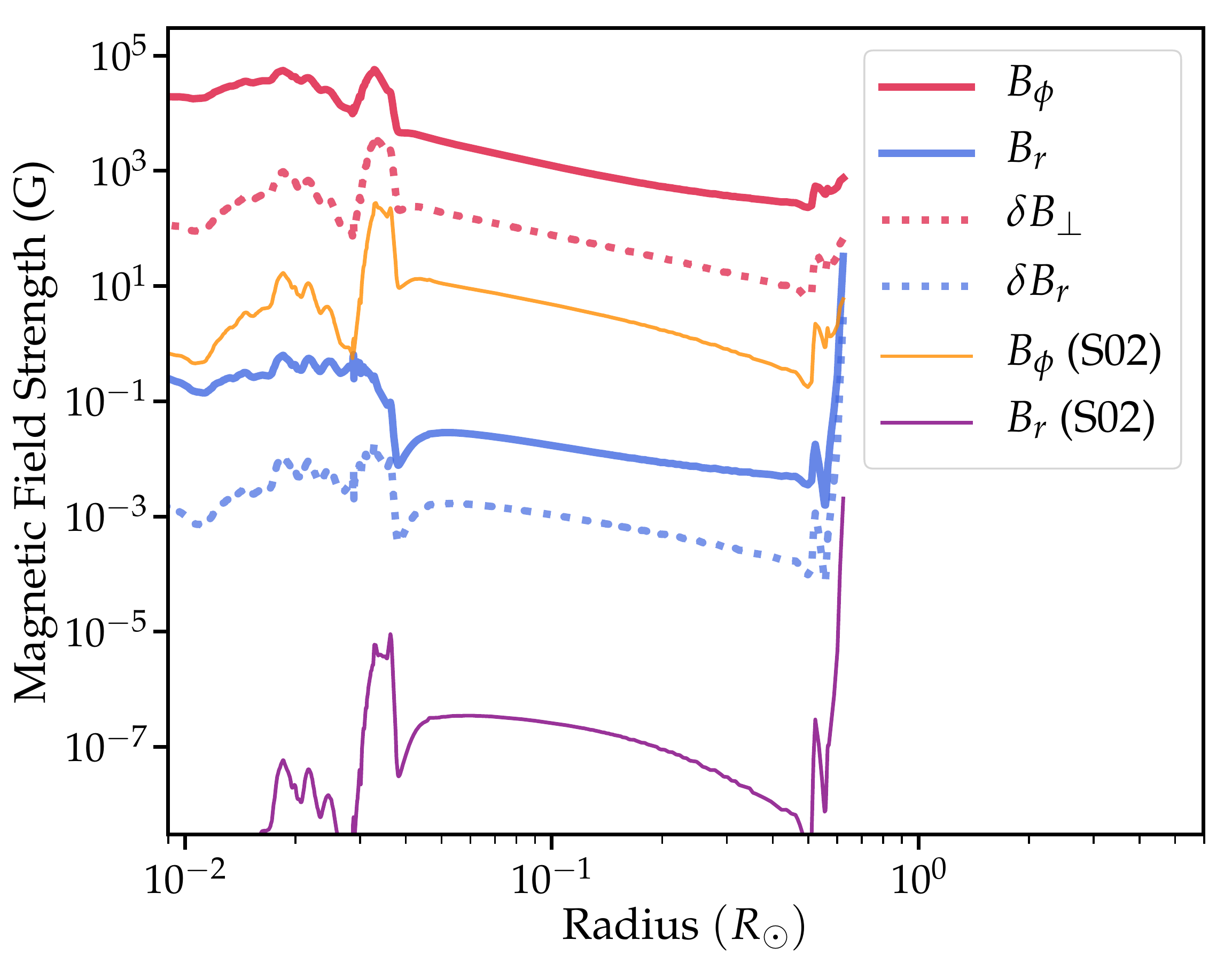}
\end{center} 
\caption{ \label{fig:bphibr} Magnetic field strengths associated with the Tayler instability in the radiative core of the same stellar model shown in Figure \ref{fig:model}. We plot the background toroidal magnetic field strength (red solid line, equation \ref{bphi1}), the mean radial magnetic field strength (blue solid line, equation \ref{br}), the typical perturbation field strength (red dashed line, equation \ref{dbperp}) and the perturbed radial field strength (blue dashed line, equation \ref{dBr}). We also plot the toroidal (orange line) and radial (purple line) field strengths from equations 21 and 23 of \citealt{spruit:02}.
}
\end{figure}

Our results have important consequences for mixing processes that depend on stellar rotation, such as meridional circulation and various shear instabilities. Mixing resulting directly from Tayler instability (equation \ref{diffr2}) is typically quite small as long as there is a composition gradient such that $\Omega/N_{\rm eff} \ll 1$, which is often the case in radiative regions that have undergone any nuclear processing. An exception to this is horizontal mixing. Unless $q \ll 1$, the horizontal circulation given by equation \ref{vperp} is more rapid than the Eddington-Sweet circulation. This helps to justify the assumption of \cite{zahn:92} that horizontal mixing is much faster than vertical circulation currents, and so supports the conclusion that vertical chemical mixing is slow relative to the vertical advection rate. More importantly, our models exhibit slower core rotation and smaller shears than previous predictions, resulting in less mixing via shear instabilities.
We thus suspect that rotational mixing has been overestimated in many previous works. To quantify this statement, more thorough calculations must be performed, incorporating AM/chemical transport via Tayler instability, meridional circulation, shear instabilities, convective overshoot, etc. The coupled effects of AM tranpsort and mixing can then be compared with abundance/rotation measurements (see e.g., \citealt{somers:16}) for stars in clusters. We hope to explore mixing effects and make detailed predictions for surface abundances in future work.

Our models assume that convection zones are nearly rigidly rotating, which may not be true for deep convective zones where asymmetric convective energy/AM fluxes may cause deeper layers of the convective envelope to rotate faster \citep{brun:09,kissin:15}. Indeed, some degree of envelope differential rotation may be necessary to explain rotation rates of horizontal branch stars \citep{sills:00}. However, envelope differential rotation does not always change our predictions for rotation in radiative cores for two reasons. First, the strong dependence of AM transport on local rotation rate, $\nu_{\rm AM} \propto \Omega^3$, causes core rotation  to converge to a rate only weakly dependent on surface rotation rate. Second, in our models we find the core rotation rate often converges to a state marginally unstable to Tayler instability such that the core rotation rate is set by equation (\ref{qmin}). Preliminary tests indicate envelope differential rotation may allow for slightly faster rotation rates of WDs, similar to the effect of decreasing the initial spin period and allowing the core to accrete more AM during the AGB.


Currently the most viable alternative to our model is that of \cite{kissin:15}, which posits rigid rotation in radiative zones enforced by magnetic torques and differential rotation in convective zones due to convective AM pumping. The clear prediction from our model is that differential rotation is mostly in the core, while the \cite{kissin:15} model predicts differential rotation confined to the evelope. Asteroseismic observations appear to disfavor envelope differential rotation \citep{dimauro:16,klion:17,dimauro:18}, though currently their ability to distinguish between the models is limited. Both models may have some tension with observations, as \cite{kissin:15} predicts rotation rates that are too slow for low-mass ($M \lesssim 1.2 M_\odot$) RGB stars, and \cite{kissin:15b} appear to predict anomalously slow rotation rates for some white dwarfs. 

A potential problem with our mechanism is that it may underpredict the scatter in observed core rotation rates, as our models converge to a similar rotation rate regardless of initial conditions. Additionally, we predict significant spin-up of red giant cores (by a factor of $\sim$2) along the lower RGB, whereas \cite{gehan:18} find no clear spin-up/spin-down as a function of evolutionary state. More work predicting spin rates for a population of stars (incorporating changes in initial spin-rate, metallicity, binarity, etc.), along with more asteroseismic measurements and a better understanding of measurement biases\footnote{We are concerned that measurement bias limits the number of core rotation measurements for stars with more rapidly rotating cores higher up the RGB, where the rotational frequency splitting becomes comparable to the mixed mode period spacing, and the asteroseismic power spectrum becomes difficult to interpret \citep{deheuvels:17}.} will help to distinguish between the competing models, though we note that differential rotation in both the convective envelope and the radiative core is possible.

Another obstacle for our model is that the radiative core must have very weak fossil fields in order for the Tayler instability to dominate AM transport. Figure \ref{fig:bphibr} plots various components of the magnetic fields in a stellar model. Of particular importance is the radial field $B_r$, which we predict to have a strength of $B_r \sim 10^{-2} \, {\rm G}$ through much of the radiative core. Recall that for the predicted value of $B_\phi$, Tayler instability cannot occur if there is a fossil field with strength greater than equation \ref{br}. If there is a fossil component of $B_r$ larger than that shown in Figure \ref{fig:bphibr}, the azimuthal component $B_\phi$ must be amplified by shear to larger field strengths before Tayler instability kicks in. By the time this occurs, the Maxwell stress $B_r B_\phi$ will be larger than predicted by our model, bringing the radiative core closer to a state of rigid rotation. So, even relatively weak fossil fields ($B_r \gtrsim 10^{-2} \, {\rm G}$) may enforce nearly rigid rotation of the radiative core. While the internal field strengths of red giants are not well known\footnote{While some red giants with suppressed dipole oscillation modes may have very strong ($B_r \gtrsim 10^5 G$) magnetic fields \citep{fuller:15,stello:16}, those whose internal rotation has been measured must have weaker fields in order for gravity waves to propagate in their cores such that the core rotation rate can be measured.}, a rigidly rotating core enforced by fossil fields would necessitate large differential rotation in the convective envelope, as advocated by \cite{kissin:15}.


\section{Conclusions}
\label{conc}

The pioneering work of \cite{spruit:02} has shown that Tayler instabilities naturally occur in stellar interiors and may dominate internal angular momentum (AM) transport. However, the saturation of the instability, and the resulting AM transport, remain poorly understood. Whereas \cite{spruit:02} posits that energy in the background field is dissipated at the instability growth rate, we argue that Tayler instability saturates via the turbulent dissipation of unstable magnetic field perturbations. Our saturation mechanism results in a smaller energy dissipation rate, such that the magnetic fields reach larger mean amplitudes. The stronger fields produce larger Maxwell stresses and more efficient AM transport. Crucially, our proposed saturation condition does not depend on the closure of a dynamo loop and thus avoids the problems pointed out by \cite{zahn:07}. Another important difference is that the minimum shear for significant AM transport (equation \ref{qmin}) is smaller than that of equation~(26) of \cite{spruit:02}, and thus Tayler instability can occur at much lower shear as long as equation (\ref{omegac}) is satisfied.

When Tayler instability operates, we find that it produces an effective AM diffusivity given by equation (\ref{AMdiff}). In our models, the resultant torque often reduces the shear to a state of marginal stability given by equation (\ref{qmin}). Implementation of our results into stellar evolution codes shows that a reasonable saturation parameter $\alpha \approx 1$ leads to core rotation rates in good agreement with asteroseismic measurements for main sequence stars, red giants, and white dwarfs across a wide range in mass. Hence, these findings may be a key step toward solving the AM transport problem within stars, and they open the door to realistic predictions of internal stellar rotation rates during phases of evolution prohibitively difficult to observe.

Our results have major implications for the core rotation rates of massive stars and their compact  progeny. Prior estimates \citep{heger:05} based on the original TS dynamo prescription predicted neutron star rotation rates of $P_{\rm NS} \sim 10 \, {\rm ms}$, somewhat faster than typical pulsar birth periods $P_{\rm NS} \gtrsim 10-50$ ms \citep{faucher:06,igoshev:13,gullon:14}. We expect our updated AM prescription to yield significantly slower NS rotation than prior predictions, and in future studies we will investigate the core rotation rates of evolving massive stars to predict the natal spin rates of neutron stars and black holes.

\section{Acknowledgments}

We thank the referee, Dr. Henk Spruit, for constructive and much-deserved criticism. We thank Peter Goldreich, Marc Pinsonneault, Matteo Cantiello, and Daniel Lecoanet for useful discussions. This research is funded in part by a Rose Hills Innovator Grant, the Gordon and Betty Moore Foundation through Grant GBMF7392, and by the National Science Foundation under Grant No. NSF PHY-1748958.

\bibliography{CoreRotBib}

\appendix

\section{Turbulent Magnetic Cascade}
\label{damp}

In this Appendix we summarize the nonlinear damping arguments of \cite{lithwick:03} and extend them to the instability discussed in this paper.
We begin with the equations of incompressible MHD in the absence of rotation or buoyancy:
\begin{align}
\label{eq:dvdt}
    \rho\partial_t \boldsymbol{v} + \rho\boldsymbol{v}\cdot\nabla\boldsymbol{v} &= -\nabla P + \rho \, \boldsymbol{v}_{\rm A}\cdot\nabla \boldsymbol{v}_{\rm A}\\
    \partial_t\boldsymbol{v}_{\rm A} + \boldsymbol{v}\cdot\nabla\boldsymbol{v}_{\rm A} &= \boldsymbol{v}_{\rm A}\cdot\nabla\boldsymbol{v}\\
    \nabla\cdot\boldsymbol{v} &= 0\\
    \nabla\cdot\boldsymbol{v}_{\rm A} &= 0.
\end{align}
The pressure $P$ is the total pressure, including both gas and magnetic contributions.
If the magnetic field is composed of a large-scale slowly-varying component and a small-scale fluctuating one we may decompose it, or equivalently the Alfv\'en velocity, as
\begin{align}
   \boldsymbol{v}_{\rm A} = \boldsymbol{v}_{{\rm A}, 0} + \delta \boldsymbol{v}_{\rm A},
\end{align}
where $\boldsymbol{v}_{{\rm A}, 0}$ is the slowly-varying background copmponent and $\delta\boldsymbol{v}_{{\rm A}}$ is the fluctuating one.
With this decomposition we define combinations of fluid velocity and magnetic field fluctuation
\begin{align}
    \boldsymbol{w}_{\pm} \equiv \boldsymbol{v} \pm \delta \boldsymbol{v}_{\rm A}.
\end{align}
The equations of motion may then be cast as
\begin{align}
\label{eq:dt}
    \partial_t \boldsymbol{w}_{\pm} \pm \boldsymbol{v}_{{\rm A}, 0} \cdot \nabla \boldsymbol{w}_{\pm} &= -\boldsymbol{w}_{\mp} \cdot \nabla \boldsymbol{w}_{\pm} - (1/\rho) \nabla P\\
    \nabla\cdot\boldsymbol{w}_{\pm} &= 0.
\end{align}
As usual in incompressible systems the pressure is not an independent degree of freedom, and may be used to ensure that the second of these equations is satisfied.
The net result is that $P$ serves to project the non-linear term in the first equation into the subspace specified by the second equation.

Studying only linear terms, equation~\eqref{eq:dt} just describes the advection of combined variations in the fluid velocity and the magnetic field.
That is,
\begin{align}
\label{eq:adv}
    \partial_t \boldsymbol{w}_{\pm} \pm \boldsymbol{v}_{{\rm A}, 0} \cdot \nabla \boldsymbol{w}_{\pm} &= 0.
\end{align}
These combined fluctuations evidently only propagate along the large-scale magnetic field.
With the addition of the non-linear terms, we see that packets of $\boldsymbol{w}_{+}$ may scatter off of those of $\boldsymbol{w}_{-}$ and vice-versa, but because the linear evolution is constrained to be along $\boldsymbol{v}_{{\rm A}, 0}$ only those packets which are bound to the same field line may scatter.

When these scattering events occur, \cite{lithwick:03} showed that they result in a bending of the field lines by an angle of order $w_{\pm} / v_{{\rm A}, 0}$.
By repeatedly bending the field-lines of a wave-packet of $\boldsymbol{w}_{\pm}$ the packet may be disrupted, such that its energy cascades non-linearly to different scales.
If the packets of $\boldsymbol{w}_{\pm}$ have wavelength $\Lambda$ parallel to $\boldsymbol{v}_{{\rm A}, 0}$ and wavelength $\lambda$ transverse to it then this disruption occurs if the field lines are displaced transversely by an amount of order $\lambda$, or by an angle of order $\lambda/\Lambda$.
If scattering events are equally likely to bend field lines in all directions the process of non-linear interactions may be described as a diffusive random walk with step size $w_{\pm} / v_{{\rm A}, 0}$, so that it takes of order
\begin{align}
    N \approx \left(\frac{\lambda}{\Lambda}\right)^2 \left(\frac{v_{{\rm A},0}}{w_\pm}\right)^2.
\end{align}
scattering events to cause a cascade.
Note that the ``strong'' regime of \cite{lithwick:03} just corresponds to the point where the above expression yields $N \la 1$.
Because each scattering event takes time of order $\Lambda/v_{{\rm A},0}$, the time-scale over which wave-packets are disrupted is
\begin{align}
    \label{tcas}
    t_{\rm cas} \approx \max(1,N) \frac{\Lambda}{v_{{\rm A},0}}.
\end{align}

The specific energy in the system, neglecting the bulk magnetic field, is of order $w_{+}^2 + w_{-}^2$.
Both $w_{+}$ and $w_{-}$ at any given scale damp due to the cascade, with packets disrupting after time-scale of order $t_{\rm cas}$.
It follows that the damping rate of the energy in the system is of orderr
\begin{align}
    \omega_{\rm damp} \approx t_{\rm cas}^{-1},
\end{align}
or equivalently, the non-linear loss rate is
\begin{align}
    \dot{E} = -\omega_{\rm damp} \rho (\delta v^2 + \delta v_{\rm A}^2).
\end{align}

We now generalize this argument to the Taylor instability.
In this case there are two additional phenomena that must be considered.
First, in a rotating system the Coriolis effect adds an acceleration to equation~\eqref{eq:dvdt}, so that
\begin{align}
\label{eq:dvdt2}
\partial_t \boldsymbol{v} + \boldsymbol{v}\cdot\nabla\boldsymbol{v} + 2 \boldsymbol{\Omega} \times \boldsymbol{v} &= -(1/\rho) \nabla P + \boldsymbol{v}_{\rm A}\cdot\nabla \boldsymbol{v}_{\rm A}.
\end{align}
The new acceleration term is not fully absorbed into the pressure gradient because it is not fully directed along the wave-vector $\boldsymbol{k}$, and hence, at least to order of magnitude, it must be kept.
Upon making the same change to equation~\eqref{eq:dt} we find
\begin{align}
    \partial_t \boldsymbol{w}_{\pm} \pm \boldsymbol{v}_{{\rm A}, 0} \cdot\nabla\boldsymbol{w}_{\pm} + \boldsymbol{\Omega}\times (\boldsymbol{w}_+ + \boldsymbol{w}_-)  \nonumber \\
     = -\boldsymbol{w}_{\mp} \cdot \nabla \boldsymbol{w}_{\pm} - (1/\rho) \nabla p.
\end{align}
Crucially, this is no longer of the form of a simple advection equation like~\eqref{eq:adv}.
Rather, packets of $w_+$ now mix with those of $w_-$ over time.
This means that the non-linear interaction, which may only act between $w_+$ and $w_-$, may act on a single wave-packet as it transitions between $w_{\pm}$.

Assuming that $\boldsymbol{w}$ is not oriented nearly-parallel to the rotation axis, the time-scale over which wave-packets transition is
\begin{align}
    t_{\rm mix} \approx \frac{w_{\pm}}{\Omega v}.
\end{align}
Recalling that we work in the limit of magnetogeostrophic balance,
\begin{align}
    w \approx |v| + |\delta v_{\rm A}| \approx \left(1 + \frac{\Omega}{\omega_{\rm A}}\right) |v| \approx \frac{\Omega}{\omega_{\rm A}} |v|,
\end{align}
so
\begin{align}
    t_{\rm mix} \approx \omega_{\rm A}^{-1}.
\end{align}
While in the mixed state, the cascade proceeds with time-scale
\begin{align}
    t_{\rm cas}^* \approx \frac{w_{\pm}}{|\boldsymbol{w}_{\mp} \cdot \nabla \boldsymbol{w}_{\pm}|} \approx \frac{\Lambda}{w_{\mp}}.
\end{align}
Analogous to the strong and weak regimes of \cite{lithwick:03}, we therefore find two regimes.
In the first, $\omega_{\rm A} t_{\rm cas}^* \ll 1$, so that scattering occurs rapidly once the wave-packets mix.
This results in an effective cascade time
\begin{align}
    t_{\rm cas} \approx \omega_{\rm A}^{-1}.
\end{align}
In the opposing limit, $\omega_{\rm A} t_{\rm cas}^* \ll 1$, scattering is slow and it makes sense to average $\boldsymbol{w}_{\mp} \cdot \nabla \boldsymbol{w}_{\pm}$ over the mixing time $\omega_{\rm A}^{-1}$.
In effect $w_+$ and $w_-$ mix quickly, a small amount of scattering occurs, and then they are unmixed again.
This repeats until the amount of scattering is of order unity, so
\begin{align}
    t_{\rm cas} \approx t_{\rm cas}^*.
\end{align}
Putting the two regimes together we find the overall effect of rotation is to reduce the cascade time-scale to
\begin{align}
    t_{\rm cas} \approx \max\left(\omega_{\rm A}^{-1}, \frac{\Lambda}{w_{\mp}}\right).
\end{align}

The above argument may also be cast in terms of new linear combinations of $\boldsymbol{v}$ and $\delta \boldsymbol{v}_{\rm A}$ which {\it do} follow an advection-like equation.
However these new linear combinations do not preserve the structure of the non-linear interaction term, and generically give rise to interactions of the form $\boldsymbol{w}_+ \cdot \nabla \boldsymbol{w}_+$, and likewise for $\boldsymbol{w}_-$.
These new self-interaction terms result, following the arguments above, in the same cascade time-cale $t_{\rm cas} \approx \Lambda/w_\pm$.

The second modification we must consider is that of buoyancy.
This works in much the same way.
We define
\begin{align}
    \boldsymbol{x} \equiv \int_0^t \boldsymbol{v}(\boldsymbol{x}(t'),t') dt'
\end{align}
as the Lagrangian displacement of a fluid element.
With this, and working in the Boussinesq limit, equation~\eqref{eq:dvdt2} becomes
\begin{align}
   \partial_t \boldsymbol{v} + \boldsymbol{v}\cdot\nabla\boldsymbol{v} &= -(1/\rho) \nabla P - \hat{r} \hat{r} \cdot \boldsymbol{x} N^2 + \boldsymbol{v}_{\rm A}\cdot\nabla \boldsymbol{v}_{\rm A},
\end{align}
where we have taken the entropy gradient to be in the radial direction and omitted the Coriolis effect for simplicity.
Once more inserting the new acceleration into equation~\eqref{eq:dt} we obtain
\begin{align}
    \partial_t \boldsymbol{w}_{\pm} \pm \boldsymbol{v}_{{\rm A}, 0} \cdot\nabla\boldsymbol{w}_{\pm} &= -\boldsymbol{w}_{\mp} \cdot \nabla \boldsymbol{w}_{\pm} - (1/\rho) \nabla P - \hat{r} \hat{r} \cdot \boldsymbol{x} N^2.
\end{align}
We may approximate the displacement as
\begin{align}
    \boldsymbol{x} \approx \frac{1}{\omega} \boldsymbol{v},
\end{align}
where $\omega$ is the linear frequency associated with any given mode.
Hence
\begin{align}
    \hat{r} \hat{r} \cdot \boldsymbol{x} N^2 \approx \hat{r} \hat{r} \cdot \boldsymbol{v} \frac{N^2}{\omega} \approx \hat{r} \hat{r} \cdot (\boldsymbol{w}_+ + \boldsymbol{w}_-) \frac{N^2}{\omega}.
\end{align}
It follows that this term, like the Coriolis one, produces mixing between $w_{\pm}$.
Because the real and imaginary parts of $\omega$ are both of order $\omega_{\rm A}^2/\Omega$, this may be written as
\begin{align}
    \hat{r} \hat{r} \cdot \boldsymbol{x} N^2 \approx \Omega \hat{r} \hat{r} \cdot (\boldsymbol{w}_+ + \boldsymbol{w}_-) \frac{N^2}{\omega_{\rm A}^2}.
\end{align}
Noting that
\begin{align}
    \boldsymbol{v}\cdot\hat{r} \approx \frac{\omega_{\rm A}}{N} v
\end{align}
we see that
\begin{align}
     \hat{r} \hat{r} \cdot \boldsymbol{x} N^2 \approx \Omega v \frac{N}{\omega_{\rm A}} \hat{r}.
\end{align}
We again use magnetogeostrophic balance to obtain
\begin{align}
    v \approx \frac{\omega_{\rm A}}{\Omega} w_{\pm}
\end{align}
so
\begin{align}
     \hat{r} \hat{r} \cdot \boldsymbol{x} N^2 \approx w_{\pm} N \hat{r}.
\end{align}
Finally we must project away the component along $\boldsymbol{k}$, because this is eliminated by the pressure gradient in geostrophic balance.
Because $k_{\perp} \approx k_r \omega_{\rm A}/N$ we find
\begin{align}
     \left(I - \hat{k}\otimes \hat{k}\right)\cdot\hat{r} \hat{r} \cdot \boldsymbol{x} N^2 \approx \omega_{\rm A} w_{\pm} .
\end{align}
It follows that this acceleration produces the same mixing and hence the same cascade rate as the Coriolis effect.

The cascade rate we have found determines the non-linear damping of the magnetic energy, so that
\begin{align}
    \frac{d}{dt}\left(\delta v_{\rm A}\right)^2 \approx -t_{\rm cas}^{-1} \left(\delta v_{\rm A}\right)^2.
\end{align}
Or phrased in terms of the linear magnetic field,
\begin{align}
    \frac{d(\delta v_{\rm A})}{dt} \approx -\frac{\delta v_{\rm A}}{t_{\rm cas}} \approx -\delta v_{\rm A} \min\left(\omega_{\rm A}, \frac{w_{\mp}}{\Lambda}\right).
\end{align}
Substituting 
\begin{align}
w_{\mp} \approx \delta v_{\rm A}
\end{align}
we find
\begin{align}
    \frac{d(\delta v_{\rm A})}{dt} \approx -\delta v_{\rm A} \min\left(\omega_{\rm A}, \frac{\delta v_{\rm A}}{\Lambda}\right),
\end{align}
such that the effetive damping rate is
\begin{align}
   \gamma \approx \min\left(\omega_{\rm A}, \frac{\delta v_{\rm A}}{\Lambda}\right)
\end{align}
For Tayler instability, the wavelength $\Lambda$ of the fastest growing modes is $\Lambda \approx 1/k_\phi \approx r/m \approx r$. In the main text, we show that $\delta v_{\rm A}/r \ll \omega_{\rm A}$, such that the effective damping rate is 
\begin{align}
    \label{gamma}
   \gamma \approx \frac{\delta v_{\rm A}}{r} \, .
\end{align}



We note that a similar result can be obtained using the heuristic argument of \cite{lithwick:03}. For Tayler instability, $\Lambda \approx \lambda \approx r$. Unlike isotropic magnetic turbulence, rotating Tayler instability is composed of magnetic perturbations that travel at group speed $v_{\rm g} \approx \omega_{\rm A}^2 r/\Omega$. Then each scattering event occurs over time scale $t_{\rm scat} \approx r/ v_{\rm g} \approx \Omega/\omega_{\rm A}^2$. Following the same argument used to derive equation \ref{tcas}, the cascade rate is then 
\begin{equation}
    t_{\rm cas}^{-1} \approx  \frac{\Omega \delta v_{\rm A}^2}{r^2 \omega_{\rm A}^2} \, .
\end{equation}
Using this result (instead of equation \ref{gamma}) in equation \ref{gammaA} yields an identical result. Additionally, we note that our saturated solution entails that
\begin{equation}
    \chi \approx \frac{ \delta v_{\rm A} \Lambda}{v_{\rm g} \lambda}  \approx 1 \,
\end{equation}
where $\chi>1$ entails strong MHD turbulence and $\chi < 1$ entails weak MHD turbulence, as defined by \cite{chandran:04}. For strong MHD turblence, the cascade rate is $t_{\rm cas} \approx  \delta v_{\rm A}/\lambda \approx \delta v_{\rm A}/r$, again equal to our result above.

\section{Energy Dissipation by Diffusively aided Field Migration}
\label{migration}

Magnetic energy in axisymmetric loops can be dissipated near the poles of the star by reconnection that connects magnetic field lines of opposing polarity. This only happens sufficiently close to the pole, where either diffusion can act across a magnetic field loop, or horizontal displacements compare to the size of the loop. Below we will show the latter length scale is larger and hence the relevant scale where dissipation occurs. The overall picture is that magnetic loops ``migrate" both poleward and equatorward due to reconnection following a Tayler displacement. After one oscillation cycle, the maximum distance a loop can migrate in the latitudinal direction is $\xi_\perp$, the horizontal displacement of a loop caused by the Tayler instability. Its value is
\beq
\label{xiperp}
\xi_\perp = \frac{\delta v_\perp}{\omega} \sim \frac{\delta v_\perp \Omega}{\omega_{\rm A}^2} \,
\eeq
where we have used the fact that the Tayler instability growth rate and oscillation frequency (i.e., the imaginary and real components of the frequency) are both $\omega \sim \omega_{\rm A}^2/\Omega$. In what follows we assume $\omega_{\rm A} < \Omega < N$ as expected in stars. 

Now, as shown in the text, the horizontal velocity is related to the perturbed magnetic field by $\delta v_\perp \sim (\omega_{\rm A}/\Omega) \delta v_{\rm A}$. Then
\beq
\label{xiperp2}
\xi_\perp \sim \frac{\delta v_{\rm A}}{\omega_{\rm A}} \, .
\eeq
We argue in the text that growth and damping of the instability are balanced when $\omega_{\rm A}^2/\Omega \sim \delta v_{\rm A}/r$. We show below that $\delta v_{\rm A}/r$ remains the relevant damping rate of the instability in spite of magnetic dissipation near the pole and any loop migration. Then we have
\beq
\label{xiperp3}
\xi_\perp \sim r \frac{\omega_{\rm A}}{\Omega} \, .
\eeq
These calculations are meant to be a mid latitudes where the cylindrical coordinate $R$ is comparable to the radial coordinate $r$. Some quantities will have different values very near the pole where $R \ll r$, but the migration time is dominated by mid latitudes where $R \sim r$, so magnetic energy can only be dissipated at the pole as fast as it migrates from mid latitudes. 

Assuming loops of azimuthal field are totally dissipated near the pole, their effective damping rate is equal to their migration rate $\gamma_{\rm mig}$. Because the loop migration is essentially a random walk process, the migration timescale is
\beq
\label{tmigrate}
t_{\rm migrate} \sim N_{\rm step}^2 t_{\rm step} \,
\eeq
where $N_{\rm step}$ is the migration length divided by a step length, and $t_{\rm step}$ is the time it takes to complete each step. The number of steps is $N_{\rm step} \sim r/\xi_\perp$. The time of each step is a magnetic diffusion time across a radial wavelength, $t_{\rm step}^{-1} \sim k_r^2 \eta$. In order for the the instability to grow, this diffusion rate must be smaller than the growth rate $\omega_{\rm A}^2/\Omega$. So we have $t_{\rm step} \gtrsim \Omega/\omega_{\rm A}^2$. Then using equations \ref{xiperp3} and \ref{tmigrate}, we have
\beq
\label{gammamig}
\gamma_{\rm migrate} = t_{\rm migrate}^{-1} \lesssim \frac{\omega_{\rm A}^4}{\Omega^3} \, . 
\eeq
This maximum migration rate will be realized when $\omega_{\rm A} \sim\omega_c$, with the critical field strength $\omega_c$ defined by equation \ref{omegac}. When $\omega_{\rm A} \sim \omega_c$, field loops can reconnect with loops of opposite oscillation phase (i.e., those separated by radial distance $\sim 1/k_r$) after $\sim 1$ oscillation cycle, such that they can migrate by a distance $\sim \xi_\perp$ each oscillation cycle. When $\omega_{\rm A} > \omega_c$, reconnection requires many oscillation cycles, the migration rate will depend on the magnetic diffusivity, and it will be smaller than equation \ref{gammamig}.

Why do we still think the instability damping rate is $\delta v_{\rm A}/r$? Let's consider whether migration of magnetic loops toward the pole can destroy them at faster rates. Assuming a loop reconnects with its neighbor after being displaced horizontally by $\xi_\perp$ at a rate $\omega_{\rm A}^2/\Omega$, the two loops have moved apart from one another at a speed $v \sim \omega_{\rm A}^2 \xi_\perp/\Omega \sim \delta v_\perp$. Then the maximum rate at which the instability can be damped due to loops migrating to the pole is
\beq
\gamma_{\rm diss} < \frac{\delta v_\perp}{r} \sim \frac{\omega_{\rm A}}{\Omega} \frac{\delta v_{\rm A}}{r} \, .
\eeq
But since $\omega_{\rm A} < \Omega$, $\gamma_{\rm diss}$ is smaller than the damping rate $\delta v_A/r$. The actual destruction rate is likely much slower due to the random walk process discussed above, and is given by equation \ref{gammamig}. This means that unstable perturbations will damp faster by weak turbulence than they will by migrating toward the pole. So the instability is still limited by weak turbulence, and setting the growth rate equal to the damping rate still implies
\beq
\frac{\omega_{\rm A}^2}{\Omega} \sim \frac{\delta v_{\rm A}}{r} \, .
\eeq

The migration rate of equation \ref{gammamig} implies that the background field, whose energy density is $\sim B_\phi^2$, is destroyed at this rate. The assosiated energy damping rate per unit volume is
\beq
\label{Edampmig}
\dot{E}_{\rm damp} \lesssim \frac{\omega_{\rm A}^4}{\Omega^3} B_\phi^2 \, .
\eeq
The turbulent energy damping discussed in the paper will operate regardless of the loop migration, and will be more important when the reconnection timescale is longer than an oscillation times scale. Hence, while damping from loop migration may be relevant when $\omega_{\rm A} \sim \omega_c$, we do not expect any of our scaling arguments or results to be altered.

\section{Effective Stratification}
\label{neffs}

As discussed in \cite{spruit:02}, the effective Brunt-V\"ais\"al\"a frequency $N_{\rm eff}$ depends on the thermal diffusion timescale across the Tayler instability lengthscale. This thermal diffusion timescale in turn depends on $N_{\rm eff}$. \cite{spruit:02} considered the limit of pure thermal/compositional stratification, but in red giant cores, both components are important. Here we derive an improved method for incorporating thermal diffusion in the general case.

Following the suggestion by \cite{spruit:02}, thermal diffusion reduces the thermal component of the effective stratification $N_{\rm T}$ by roughly
\beq
N_{\rm T,eff}^2 = \frac{N_{\rm T}^2}{1 + k^2 \chi/\omega} \, ,
\eeq
where $k$ is the instability wavenumber, $\chi$ is the thermal diffusivity, and $\omega$ is the instability timescale. The fastest growing modes have $r k \sim N_{\rm eff}/\omega_{\rm A}$ and $\omega = \omega_{\rm A}^2/\Omega$. Using the saturated value of $\omega_{\rm A}$ from equation \ref{bphi}, we have
\beq
N_{\rm T}^2= \bigg(1 + \frac{\omega_t}{\Omega} \bigg) N_{\rm T,eff}^2 \, .
\eeq
The effective stratification is $N_{\rm eff}^2 = N_{\rm T,eff}^2 + N_\mu^2$, where $N_\mu$ is the compositional component of the stratification. Substituting for $N_{\rm T,eff}$, we find 
\beq
\label{neffeq}
N_{\rm eff}^2 - N^2 + (N_{\rm eff}^2 - N_\mu^2)\frac{\omega_t}{\Omega} = 0  .
\eeq
Equation \ref{neffeq} can be solved for the appropriate value of $N_{\rm eff}$ given a stellar structure. Inspection reveals that it reduces in the appropriate limits. When $\chi \rightarrow 0$, we recover $N_{\rm eff} = N$. When $\chi \rightarrow \infty$, we recover $N_{\rm eff} = N_\mu$. And when $N_\mu \rightarrow 0$ and thermal diffusion is large, $N_{\rm eff} \ll N$ and we find $N_{\rm eff} = (r^6 q^4 \alpha^{12} \Omega^{13} N^6/\chi^3)^{1/16}$. We have not yet implemented numerical solutions of equation \ref{neffeq} into our MESA routines, but we plan to do this in future work.

\section{MESA Model Inlists}
\label{appendixmodel}

We use the MESA stellar evolution code \cite{paxton:11,paxton:13,paxton:15,paxton:18} version 10108 to generate our stellar models. The inlist for our models is as follows:

\begin{verbatim}
&star_job
            
      pgstar_flag = .true.

      new_rotation_flag = .true.
      change_rotation_flag = .true.
      change_initial_rotation_flag = .true.

      new_omega = 3.64e-5
      set_initial_omega = .true.

/ ! end of star_job namelist

&controls

      !------------------------  MAIN
       
      initial_mass = 1.6
      initial_z = 0.02
      use_Type2_opacities = .true.
      Zbase = 2.d-2 

      set_min_D_mix = .true.
      min_D_mix = 1d1
 
      mesh_delta_coeff = 0.7
      varcontrol_target = 0.7d-3

      predictive_mix(1) = .true.
      predictive_superad_thresh(1) = 0.005
      predictive_avoid_reversal(1) = 'he4'
      predictive_zone_type(1) = 'any'
      predictive_zone_loc(1) = 'core'
      predictive_bdy_loc(1) = 'top'

      dX_div_X_limit_min_X = 1d-4
      dX_div_X_limit = 5d-1
      dX_nuc_drop_min_X_limit = 1d-4
      dX_nuc_drop_limit = 1d-2
      

      !---------------------  Rotation

      am_nu_ST_factor = 0
      use_other_am_mixing = .true.

      am_time_average = .true.
      premix_omega = .true.
      recalc_mixing_info_each_substep = .true.
      am_nu_factor = 1
      am_nu_non_rotation_factor = 1d0
      am_nu_visc_factor = 0.333
      angsml = 0.0

      !-------------------------  WIND

      cool_wind_RGB_scheme = 'Reimers'
      cool_wind_AGB_scheme = 'Blocker'
      RGB_to_AGB_wind_switch = 1d-4
      Reimers_scaling_factor = 0.2
      Blocker_scaling_factor = 0.5
      use_accreted_material_j = .true.
      accreted_material_j = 0

      !-------------------  OVERSHOOTING
 
      overshoot_f_above_nonburn_core = 0.015
      overshoot_f0_above_nonburn_core = 0.005
      overshoot_f_above_nonburn_shell = 0.015
      overshoot_f0_above_nonburn_shell = 0.005
      overshoot_f_below_nonburn_shell = 0.015
      overshoot_f0_below_nonburn_shell = 0.005

      overshoot_f_above_burn_h_core = 0.015
      overshoot_f0_above_burn_h_core = 0.005
      overshoot_f_above_burn_h_shell = 0.015
      overshoot_f0_above_burn_h_shell = 0.005
      overshoot_f_below_burn_h_shell = 0.015
      overshoot_f0_below_burn_h_shell = 0.005

      overshoot_f_above_burn_he_core = 0.015
      overshoot_f0_above_burn_he_core = 0.005
      overshoot_f_above_burn_he_shell = 0.015
      overshoot_f0_above_burn_he_shell = 0.005
      overshoot_f_below_burn_he_shell = 0.015
      overshoot_f0_below_burn_he_shell = 0.005

/ ! end of controls namelist
\end{verbatim}

Some important controls include the use of predictive mixing to help mitigate ``breathing pulses" in the size of the convective helium-burning core during the clump. Additionally, the use of \verb|am_time_average|, \verb|premix_omega|, and \verb|recalc_mixing_info_each_substep| help reduce numerical artifacts related to AM transport. Smoothing the shear and AM diffusivity (see next section) also help reduce these numerical instabilities. The artifacts arise because large MESA timesteps can cause AM transport to artificially create step-like features in the stellar rotation profile. The steps arise where AM transport in some grid cells is slightly more efficient than neighboring grid cells due to the discrete grid size and inaccurate numerical derivatives. A large time step will cause the rotation profile to flatten in grid cells with larger AM diffusivity, and steepen in neighboring grid cells with smaller AM diffusivity. The controls above help mitigate these effects, but in some cases enforcing smaller time steps may be useful.

The initial masses and rotation rates are adjusted as described in the text. In some models  we adjust the wind scaling factors on the AGB in order to avoid late helium flashes. We also enable MLT++ to evolve more massive stars from the AGB to the WD cooling track:
\begin{verbatim}
    okay_to_reduce_gradT_excess = .true.
    gradT_excess_max_change = 1d-2
\end{verbatim}
and in some cases we remove the last few hundredths of a solar mass of the hydrogen envelope using
\begin{verbatim}
    remove_H_wind_mdot = 1d-4
    remove_H_wind_H_mass_limit = 1d-5
\end{verbatim}
We have performed some basic resolution testing to verify our results are very insensitive to the model's grid resolution and time stepping.

\newpage
\onecolumn

\subsection{Implementation of Angular Momentum Transport}

Our \verb|run_star_extras.f| code for implementation of AM transport in our MESA models is as follows:

\begin{verbatim}
subroutine TSF(id, ierr)

  integer, intent(in) :: id
  integer, intent(out) :: ierr
  type (star_info), pointer :: s
  integer :: k,j,op_err,nsmooth,nsmootham
  real(dp) :: alpha,shearsmooth,nu_tsf,nu_tsf_t,omegac,omegag,omegaa,omegat
  real(dp) :: difft,diffm,brunts,bruntsn2,logamnuomega,alphaq

  call star_ptr(id,s,ierr)
  if (ierr /= 0) return

  alpha=1d0
  nsmooth=5
  nsmootham=nsmooth-3
  shearsmooth=1d-30
  op_err = 0

  !Calculate shear at each zone, then calculate TSF torque
  do k=nsmooth+1,s% nz-(nsmooth+1)

    nu_tsf=1d-30
    nu_tsf_t=1d-30
    !Calculate smoothed shear, q= dlnOmega/dlnr
    shearsmooth = s% omega_shear(k)/(2.*nsmooth+1.)
    do j=1,nsmooth
      shearsmooth = shearsmooth + (1./(2.*nsmooth+1.))*( s% omega_shear(k-j) + s% omega_shear(k+j) )
    end do

    !Magnetic diffusivity
    diffm =  diffmag(s% rho(k),s% T(k),s% abar(k),s% zbar(k),op_err) 
    !Thermal diffusivity
    difft = 16d0*5.67d-5*(s% T(k))**3/(3d0*s% opacity(k)*(s% rho(k))**2*s% Cv(k)) 
    !Alfven frequency at saturation
    omegaa = s% omega(k)*(shearsmooth*s% omega(k)/sqrt(abs(s% brunt_N2(k))))**(1./3.) 
    !Thermal damping rate assuming adiabatic instability
    omegat = difft*pow2(sqrt(abs(s% brunt_N2(k)))/(omegaa*s% r(k)))
    !Suppress thermal part of brunt 
    brunts = sqrt(abs( s% brunt_N2_composition_term(k) + 
       (s% brunt_N2(k)-s% brunt_N2_composition_term(k))/(1d0 + omegat/omegaa) )) 
    !Effective brunt for isothermal instability
    bruntsn2 = sqrt(abs( s% brunt_N2_composition_term(k) + 
       (s% brunt_N2(k)-s% brunt_N2_composition_term(k))*min(1d0,diffm/difft) )) 
    !Choose max between suppressed brunt and isothermal brunt
    brunts = max(brunts,bruntsn2) 
    !Don't let Brunt be smaller than omega
    brunts = max(s% omega(k),brunts) 
    !Recalculate omegaa
    omegaa = s% omega(k)*abs(shearsmooth*s% omega(k)/brunts)**(1./3.) 

    !Calculate nu_TSF
    if (s% brunt_N2(k) > 0.) then
      if (pow2(brunts) > 2.*pow2(shearsmooth)*pow2(s% omega(k))) then
        !Critical field strength
        omegac = 1d0*s% omega(k)*((brunts/s% omega(k))**0.5)*(diffm/(pow2(s% r(k))*s% omega(k)))**0.25  
        !Suppress AM transport if omega_a<omega_c
        nu_tsf = 5d-1+5d-1*tanh(5d0*log(alpha*omegaa/omegac)) 
        !nu_omega for revised Tayler instability
        nu_tsf = nu_tsf*alpha**3*s% omega(k)*pow2(s% r(k))*(s% omega(k)/brunts)**2 
      end if
      ! Add TSF enabled by thermal diffusion
      if (pow2(brunts) < 2.*pow2(shearsmooth)*pow2(s% omega(k))) then
        nu_tsf_t = alpha*abs(shearsmooth)*s% omega(k)*pow2(s% r(k))
      end if
      s% am_nu_omega(k) = s% am_nu_omega(k) + max(nu_tsf,nu_tsf_t) + 1d-1
    end if

  end do


	 
  !Smooth nu_omega
  logamnuomega=-3d1
  do k=nsmootham+1,s% nz-(nsmootham+1)
  
    !Don't smooth convective diffusivity into non-convective zones
    if (s% mixing_type(k)==1) then
       s% am_nu_omega(k) = s% am_nu_omega(k)
       !Smooth zones if not including a convective zone
    else
      logamnuomega = log10(s% am_nu_omega(k))/(2.*nsmootham+1.)
    end if 
    
    do j=1,nsmootham
      !Don't smooth convective diffusivity into non-convective zones
      if (s% mixing_type(k-j)<3.5) then
        logamnuomega = log10(s% am_nu_omega(k))
        !Smooth zones if not including a convective zone
      else 
        logamnuomega = logamnuomega + (1./(2.*nsmootham+1.))*log10(s% am_nu_omega(k-j)) 
      end if
    end do
    
    do j=1,nsmootham
      !Don't smooth convective diffusivity into non-convective zones
      if (s% mixing_type(k+j)<3.5) then
        logamnuomega = logamnuomega
        !Smooth zones if not including a convective zone
      else 
        logamnuomega = logamnuomega + (1./(2.*nsmootham+1.))*log10(s% am_nu_omega(k+j))
      end if
    end do
    
    s% am_nu_omega(k) = 10.**logamnuomega
  end do

  !Values near inner boundary
  do k=s% nz-nsmootham,s% nz
    s% am_nu_omega(k) = s% am_nu_omega(k-1)
  end do

  !Values near outer boundary
  do k=nsmootham,1
    s% am_nu_omega(k) = s% am_nu_omega(k-1)
  end do

end subroutine TSF
\end{verbatim}

These controls work well for our models, but we caution that they may not work well in different situations. For instance, we estimate $N_{\rm eff}$ in a way which is accurate for our models but may be problematic in some stars. Magnetic diffusivity is calculated via the modules included in MESA's default implementation for TS torques. To disable our AM transport prescription when $\omega_{\rm A} \! < \! \omega_c$, we use a \verb|tanh| function to smoothly transition from no torque at $\omega_{\rm A} \! < \! \omega_c$ to full torque at $\omega_{\rm A} \! > \! \omega_c$. Additionally, we smooth the dimensionless shear by 5 grid cells on each side, and we smooth the AM diffusivity by 32 grid cells on each side. In our models, this level of smoothing helps suppress numerical instabilities but does not strongly affect the evolution because larger smoothing lengths deliver nearly identical results.

\end{document}